\documentclass[a4paper,11pt]{article}
\pdfoutput=1 % if your are submitting a pdflatex (i.e. if you have
\usepackage[utf8]{inputenc}
\usepackage[T1]{fontenc}
\usepackage{graphicx}
\usepackage{amsmath}
\usepackage{empheq}
\usepackage{amsfonts}
\usepackage{jcappub} % for details on the use of the package, please
\usepackage{epsfig}
\usepackage{epstopdf}
\input{epsf.sty}
\usepackage{epsf}
\usepackage{amssymb}
\usepackage{caption}
\usepackage{subcaption}
\usepackage{tensor}
\usepackage{bm}
\usepackage{color}
\usepackage{dsfont}
\usepackage{wasysym}
\usepackage{amsthm}
\usepackage{hyperref}
\usepackage{mathrsfs}
\usepackage{epsfig,bbm}
\usepackage{xcolor}
\usepackage{mathtools}
\usepackage{empheq}
\usepackage[normalem]{ulem}

\newcommand*\widefbox[1]{\fbox{\hspace{2em}#1\hspace{2em}}}
\def\be{\begin{equation}}
\def\ee{\end{equation}}
\def\ba{\begin{eqnarray}}
\def\ea{\end{eqnarray}}
\def\bea{\begin{eqnarray}}
\def\eea{\end{eqnarray}}
\include{mydefs}
\newcommand{\bal}{\begin{eqnarray}}
\newcommand{\eal}{\end{eqnarray}}

\newcommand{\no}{\nonumber}

\begin{document}
\begin{flushright}
	PI/UAN-2017-607FT
\end{flushright}

\title{de Sitter symmetries and inflationary correlators in parity violating scalar-vector models }

\author[a, 1]{Juan P. Beltr\'an Almeida,\note{Corresponding author.}}
\author[b]{Josu\'e Motoa-Manzano,}
\author[b]{C\'esar A. Valenzuela-Toledo,}

\affiliation[a]{Departamento de F\'isica, Universidad Antonio Nari\~no, \\ Cra 3 Este \# 47A-15, Bogot\'a DC, Colombia}
\affiliation[b]{Departamento de F\'isica, Universidad del Valle, \\
Cll. 13 \# 100 - 0 A.A. 24360, Cali, Colombia}

% e-mail addresses: one for each author, in the same order as the authors
\emailAdd{ juanpbeltran@uan.edu.co}
\emailAdd{cesar.valenzuela@correounivalle.edu.co}
\emailAdd{josue.motoa@correounivalle.edu.co}
%\emailAdd{fourth@one.univ}

\vskip 0.5cm 

\abstract{ In this paper we use conformal field theory techniques to constrain the form of the correlations functions of an inflationary scalar-vector model described by the interaction term $f_1(\phi)F_{\mu \nu}F^{\mu \nu} + f_2(\phi)\tilde{F}_{\mu \nu}F^{\mu \nu}$. We use the fact that the conformal group is the relevant symmetry group acting on super horizon scales in an inflationary de Sitter background. As a result, we find that super horizon conformal symmetry, constraints the form of the coupling functions $f_1, f_2$ to be homogeneous functions of the same degree. We derive the general form of the correlators involving scalar and vector perturbations in this model and determine its squeezed limit scaling behaviour for super horizon scales. The approach followed here is useful to constraint the shape of scalar-vector correlators, and our results agree with recent literature on the subject, but don't allow us to determine amplitude factors of the correlators. 
}
 
\keywords{Statistical anisotropy,  parity violation, non-gaussianity, gravitational waves}

\maketitle

%%%%%%%%%%%%%%%%%%%%%%%%%%%%
%%%%%%%%%%%%%%%%%%%%%%%%%%%%
\section{Introduction}
%%%%%%%%%%%%%%%%%%%%%%%%%%%%
%%%%%%%%%%%%%%%%%%%%%%%%%%%%
It is well known that symmetries impose constraints on the dynamics and on the observables represented by the correlation functions of a physical system. During inflation, the relevant symmetry group of the space time is the de Sitter group, so, we expect that this group gives us hints about the structure of the correlation functions of the different fields present during the inflationary epoch  \cite{Antoniadis:1996dj,Strominger:2001pn,Maldacena:2011nz, Antoniadis:2011ib,Coriano:2013jba}. Over the years, several studies had explored the roll of the de Sitter group during inflation \cite{Antoniadis:1996dj,Strominger:2001pn, Maldacena:2011nz,Antoniadis:2011ib,Creminelli:2011mw,Creminelli:2012ed,Kehagias:2012pd,Kehagias:2012td, Coriano:2013jba, Biagetti:2013qqa}. A crucial point in this study is the fact that, for super horizon scales, the de Sitter symmetry group approaches the behavior of the conformal group in one dimension below, this is, the de Sitter group acting on four dimensional de Sitter space, acts as the conformal group in the euclidean three dimensional space. 
This fact has been used  to characterize the corelators of primordial inflationary perturbations 
\cite{Maldacena:2011nz,Creminelli:2011mw,Creminelli:2012ed,Kehagias:2012pd,Kehagias:2012td,Biagetti:2013qqa}. In particular, when the primordial perturbations are sourced by light fields other than the inflaton, the super horizon perturbations become
conformally invariant and then, their correlation functions  are constrained by this symmetry, allowing us to fix the shape of the inflationary correlators \cite{Antoniadis:1996dj,Maldacena:2011nz,Antoniadis:2011ib,Creminelli:2011mw,Creminelli:2012ed,Kehagias:2012pd,Kehagias:2012td} (other, related  uses of symmetry arguments in this context can be found in  {\it e.g.} \cite{Dimastrogiovanni:2014ina,Dimastrogiovanni:2015pla,Marcori:2016oyn}). It has also been shown that this reasoning hold even when perturbations are coming from vector fields \cite{Biagetti:2013qqa}. Likewise, this tool has  been used to establish a set of consistency relations that involve the correlations functions of matter in the process of large scales structure formation of the universe \cite{Kehagias:2013yd,Peloso:2013spa,Creminelli:2013mca}.

On the other hand, observational data shows us that there are some evidences that support the existence of small departures from the standard cosmological scenario, {{\it i.e.}  the cosmological  perturbations  exhibit tiny departures  from scale invariance, gaussianity, isotropy and homogeneity among other deviations \cite{Ade:2015hxq,Ade:2015ava}}. 
This observational evidence is usually gathered in a set of observational signatures that are called {\it cosmic anomalies} \cite{Perivolaropoulos:2014lua,Ade:2015hxq,Ade:2015ava} (see for example Ref \cite{Perivolaropoulos:2014lua} for a  description of each one). Although this anomalies could be due to some sort of statistical fluke, a minimal signal of their existence force us to modify  the simplest models, usually builded with scalar fields. A natural option of the above modifications, is to include vector fields in the inflationary dynamics, because they naturally could be responsible of several interesting possibilities such as breaking of statistical isotropy, parity violating patterns, the origin of primordial magnetic fields, or even support an inflationary period, among others\footnote{It is well know that models where vector fields support the inflationary expansion are usually ruled out due to stability problems or observational constraints (see for instance \cite{Himmetoglu08a,Himmetoglu:2008hx,Himmetoglu:2009qi,Namba:2013kia,Adshead:2013nka}). However, in some cases, these problems can be solved by introducing non canonical kinetic couplings \cite{Adshead:2016omu,Adshead:2017hnc}.}  (see for instance \citep{Dima10a,Maleknejad12,Soda12} and references therein). These new features are usually confronted with observations because they can become in an interesting discriminating tests of the proposed models \cite{Ackerman:2007nb,Namba:2013kia,Adshead:2013nka,Kim:2013gka}.

Some inflationary models consider a coupled scalar-vector system, where the scalar field is responsible for the inflationary expansion, and the vector fields can contribute to the primordial curvature perturbation and, at the same time, could explain some of the inflationary anomalies \cite{Dima10a,Maleknejad12,Soda12}. In this paper we follow the approach proposed in Ref. \cite{Biagetti:2013qqa} to study the conformal symmetry constraints of the coupled scalar-vector model $f(\phi)F^2$ \cite{Ratra:1991bn,Yokoyama08,Watanabe:2009ct,Watanabe10,Caldwell:2011ra,Bartolo12,Soda12}. We extend the symmetry analysis in \cite{Biagetti:2013qqa} to the model with a coupling term of the form
 \be\label{SVF}
f_{1}(\phi)F^{\mu \nu}F_{\mu \nu} + f_{2}(\phi)F^{\mu \nu}\tilde{F}_{\mu \nu}\;,
\ee
which explicitly introduces parity violation on the action. We find that this model respects three dimensional conformal invariance on super horizon scales only when the coupling functions are proportional $f_2 = \gamma f_1$\footnote{This result is a consequence of the exact conformal symmetry, therefore it is only valid in pure de Sitter space-time.}. In this case we calculate the conformal weight of the fields in the asymptotic future of de Sitter space, which allows us to characterize the correlators of the conformal field theory at the boundary which we identify with the super horizon scales regime.  Our main goal is to determine the form of the two and three point correlation functions on this model and see what new features raise by breaking rotational invariance and parity symmetry \cite{Caprini:2014mja,Caprini:2017vnn,Bartolo:2015dga}. 

The paper is organized as follows, in section \ref{Gen} we briefly review the basics of de Sitter symmetry group and its realisation as the conformal group in three dimensional euclidean space. In section \ref{VFmodel} we describe the dynamics of the system and analyse its asymptotic super horizon behaviour. Later, in section \ref{CorF} we calculate the most general form of the correlation functions in the super horizon limit that are allowed by  conformal symmetries. We end in section \ref{Con} with the discussion of our results.  

%%%%%%%%%%%%%%%%%%%%%%%%%%%%
%%%%%%%%%%%%%%%%%%%%%%%%%%%%
\section{Asymptotic symmetries of de Sitter space}\label{Gen}
%%%%%%%%%%%%%%%%%%%%%%%%%%%%
%%%%%%%%%%%%%%%%%%%%%%%%%%%%

\subsection{Isometries of the de Sitter spacetime}
%%%%%%%%%%%%%%%%%%%%%%%%%%%%
In 4D conformal planar coordinates, the line element of de Sitter spacetime is written as 
\be
\label{ds}
ds^2=a(\tau)^2\left(-d\tau^2+d\vec{x}^2\right)=\frac{1}{\left(-H\tau\right)^2}\left(-d\tau^2+d\vec{x}^2\right)\;,
\ee
where $a$ is the scale factor, $H$ is the Hubble parameter and  $\tau$ is the {\it conformal time}. In this coordinate system it is easy to see that the line element (\ref{ds}) is invariant under the following transformations (isometries of the de Sitter metric):\\
$i)$ Spacial translations and rotations on $\tau = constant$ slicings:
\be
 x'_i=a_i+R_{ij}x_j.\label{ltz}
\ee
$ii)$ Spacetime dilatation:  
\be\label{dilation}
x'^\mu=\lambda x^\mu \rightarrow \left\{\begin{array}{l}
x'_i=\lambda x_i, \\ \\
\tau'=\lambda \tau\;.
\end{array}\right.
\ee 
$iii)$ Special conformal transformations (SCT):
\be\label{CST}
x'^\mu=\frac{x^\mu+b^\mu  {x}^2 }{1+2\vec{b}\cdot\vec{x}+
b^2  {x}^2 } \rightarrow \left\{\begin{array}{l}
x'_i=\frac{x_i+b_i\left(-\tau^2+\vec{x}^2\right)}{1+2\vec{b}\cdot\vec{x}+
b^2\left(-\tau^2+\vec{x}^2\right)}\;, \\ \\
 \tau'=\frac{\tau}{1+2\vec{b}\cdot \vec{x}+b^2\left(-\tau^2+\vec{x}^2\right)}\;.
\end{array}\right.
\ee 
In the above expressions $a_i$ is a 3-dimensional constant vector, $R_{ij}$ is an $O(3)$ rotation matrix satisfying $R_{ik}R^{kj}=\delta_i^j$\footnote{Notice that the $O(3)$ group is consistent with the presence of inversions $\vec{x} \rightarrow -\vec{x}$ which are allowed since the term $F\tilde{F}$ allows for symmetry under inversions.}, $\lambda$ is a constant factor, $b^\mu = (0, b_i)$ is a three dimensional vector which generates the transformation and $\eta_{\mu\nu}$ is the Minkowski metric. We use the notation $x^2 = \eta_{\mu\nu} x^\mu x^{\nu}$, $\eta_{\mu\nu}= {\rm diag}(-1, 1, 1,1)$, $\vec{x}^2 = \delta_{ij} x^ix^j$ and $\vec{x}\cdot\vec{b} = \delta_{ij} b^ix^j$. Together, the transformations (\ref{ltz})-(\ref{CST}) form a set of ten symmetry generators, which is what we expect from a maximally symmetric four dimensional spacetime. 

%%%%%%%%%%%%%%%%%%%%%%%%%%%%
\subsection{Conformal group basics and relation with de Sitter group  symmetries }
%%%%%%%%%%%%%%%%%%%%%%%%%%%%
In an Euclidean space, the conformal group is defined as the group of transformations that leave invariant the metric up to a factor, this is,
\be \label{confg}
g_{\mu \nu}(x) \rightarrow  g'_{\mu \nu}(x') = \Omega^2(x)g_{\mu \nu}(x),
\ee
where $\Omega(x)$ is an arbitrary function of the coordinates. It is easy to see that the transformations (\ref{ltz})-(\ref{CST})  act as conformal transformations on $\mathbb{R}^3$ for $ |\tau| \ll |\vec{x}|$ with the euclidean metric $g_{ij} = \delta_{ij}$. According with (\ref{confg}), the corresponding conformal factor $\Omega^2$ for the transformations are:
\ba \label{tr3d}
x'_i = a_i+R_{ij} x_j  & \rightarrow & \Omega^2 = 1,\\ \label{dil3d}
x'_i = \lambda x_i  & \rightarrow & \Omega^2 = \lambda^{-2}, \\ \label{sct3d}
x'_i =\frac{x_i+b_i \vec{x}^2 }{1+2\vec{b}\cdot\vec{x}+
b^2\left(\vec{x}^2\right)}  & \rightarrow & \Omega^2 = (1+2\vec{b} \cdot \vec{x} + b^2 \vec{x}^2 )^2.
\ea
So, we see that in the asymptotic region $-H\tau\rightarrow 0$, the de Sitter group  ${\rm SO}(4, 1)$ is isomorphic to the conformal group in three dimensional euclidean space. In other words, the symmetries (\ref{tr3d})-(\ref{sct3d}) are the {\it asymptotic symmetry group} of the  boundary region of de Sitter space. In the context of inflation, the asymptotic region correspond to the super Hubble scales where the primordial perturbations become classical to later evolve according to the gravitational instability \cite{Bernardeau:2001qr}. The main idea is that, for perturbations with scales well outside of the horizon, with $-k\tau \rightarrow 0$, the form of the correlators can be fixed by conformal invariance \cite{Antoniadis:1996dj,Antoniadis:2011ib,Creminelli:2012ed,Kehagias:2012pd,Kehagias:2012td,Biagetti:2013qqa}. 
Finally, we recall that a conformal tensor field $T_{i_1...i_r}(x)$ is defined in such a way that, under conformal transformations, transforms as: 
 \ba\label{tl}
T_{i_1...i_r}(x) \rightarrow T'_{i_1...i_r}(x')=\left|\det\left(\frac{\partial x'^l}{\partial x^k}\right)\right|^{\frac{r-\Delta_T}{d}}\frac{\partial x^{j_1}}{\partial x'^{i_1}}...\frac{\partial x^{j_r}}{\partial x'^{i_r}} T_{j_1...j_r}(x)\;,
\ea
where $r$ is the rank of the tensor, $d$ is the space dimension  and $\Delta_T$ is the {\it conformal weight} or  {\it conformal dimension} of the field. By using this transformation, we can find the conformal weight $\Delta_T$ of 
the fields involved in the theory, 
which are the necessary elements to describe the structure and scaling of the correlators of the theory. For a detailed review of the basics about conformal field theory see \cite{Ginsparg:1988ui}. 
%%%%%%%%%%%%%%%%%%%%%%%%%%%%
%%%%%%%%%%%%%%%%%%%%%%%%%%%%
\section{A parity violating scalar-vector inflationary model}\label{VFmodel}
%%%%%%%%%%%%%%%%%%%%%%%%%%%%
%%%%%%%%%%%%%%%%%%%%%%%%%%%%
As said before, we are interested in inflationary models which include vector fields as a source for statistical deviations from isotropy and parity symmetry. We consider a scenario with a scalar field driving the inflationary expansion, and a vector field, which is partly responsible fro the generation of the primordial curvature perturbation and is a source of statistical anisotropies. Several models of this type has been studied in the literature \citep{Ratra:1991bn,Dimopoulos09a,Dimopoulos09vu,Watanabe10,Dimopoulos10xq,Barnaby:2011vw,Sorbo:2011rz,Dimopoulos:2012av,Anber:2012du,Bartolo12,Lyth13,Shiraishi:2013kxa,Cook:2013xea,Nurmi:2013gpa,Caprini:2014mja,Chen:2014eua,Bartolo:2015dga,Namba:2015gja, Valenzuela2016}. The  interest for those models has been diverse and it has evolved over the years. It is worthwhile to mention the Ratra model \cite{Ratra:1991bn}, which has a rich and interesting  phenomenology  in the context of primordial magnetogenesis scenario \cite{Caldwell:2011ra,Motta:2012rn,Jain:2012ga,Jain:2012vm,Durrer:2013pga}   and later used as a source of sizable statistical anisotropies which doesn't dilute with the inflationary evolution \cite{Yokoyama08, Watanabe:2009ct, Watanabe10, Soda12, Bartolo12}.  Aside of this, there is a rich literature about models whit an axion/pseudoscalar coupling of the form $\phi F\tilde{F}$ which allows for interesting phenomenology and source of non-vacuum gravitational waves, chiral gravitational waves, large tensor non-gaussianities among other interesting effects \cite{Sorbo:2011rz, Anber:2012du, Cook:2013xea, Caprini:2014mja, Namba:2015gja}.  In this paper we consider a mixture of this two scenarios which can be encompassed in the following model: 
\ba\label{pvgm} 
S &=& \int d^{4}x \sqrt{-g}\left[  \frac{M_{pl}^{2}}{2} R - \frac{1}{2} \partial_{\mu}\phi \partial^{\mu}\phi - V(\phi)  -\frac{1}{4}f_{1}(\phi)F^{\mu \nu}F_{\mu \nu} - \frac{1}{4}f_{2}(\phi)F^{\mu \nu}\tilde{F}_{\mu \nu} \right]
\ea
where $f_1$ and $f_2$ are functions which depend only of the scalar field and $\tilde{F}^{\mu \nu}$ is the Hodge dual of the field strength $F_{\mu \nu}$ which is defined by $\tilde{F}_{\mu \nu} = \frac{1}{ 2\sqrt{-g}} \varepsilon_{\mu \nu \alpha \beta} F^{\alpha \beta}$.
This model incorporate the features of the Ratra model  $f_{1}(\phi)F^{\mu \nu}F_{\mu \nu}$  with all the phenomenology carried with the pseudoscalar term $f_{2}(\phi)F^{\mu \nu}\tilde{F}_{\mu \nu}$. The interaction term 
\ba \label{ccsv}
S_{\rm{\phi A}} = -\frac{1}{4}\int d^{4}x \sqrt{-g}   \left[ f_{1}(\phi) F^{\mu \nu}F_{\mu \nu}  +  f_{2}(\phi)\tilde{F}^{\mu \nu}F_{\mu \nu} \right],\;
\ea
is responsible for the introduction of non diluting anisotropies and for sourcing parity violating signatures on the CMB correlators. Although this term represent a non-renormalizable, non-canonical interacting kinetic term between the scalar and vector fields, it has been seen that this terms represent a model which is stable, in the sense that the Hamiltonian of the theory is  bounded by below when $f_1$ is positive and the theory is causal, since the equations of motion are hyperbolic  \cite{Fleury14qfa}. Further generalizations could introduce multiple scalar and vector fields but all the main features can be captured in this single scalar-vector interaction.

%%%%%%%%%%%%%%%%%%%%%%%%%%%%
\subsection{Equations of motion and asymptotic behavior}\label{asymptotic_solutions}
%%%%%%%%%%%%%%%%%%%%%%%%%%%%
Now we focus on the dynamics of the fields propagating  on de Sitter space. Before study the vector field model, let us recall some basics about the single massive scalar field on de Sitter. The action for this case reads
\be
S_\phi=-\frac{1}{2}\int d^4x \sqrt{-g} \left( \nabla^{\mu}\phi \nabla_{\mu}\phi + m^2 \phi^2\right),\label{aces1}
\ee
where $m$ is the mass of the field. In the coordinates (\ref{ds}), the equation of motion becomes
\be\label{eqphi}
 {\phi}''-\frac{2}{\tau} {\phi}'-\nabla^2 \phi+\frac{m^2\phi}{H^2\tau^2}=0,
\ee
and its asymptotic future solution ($-H\tau \rightarrow 0$)  is (see {\it e.g.} \cite{Riotto:2002yw}):
\be
\phi(\vec{x}, \tau) \approx  C_{+} \tau^{h_ {+}} \sigma_+ (\vec{x}) + C_{-} \tau^{h_ {-}} \sigma_{-} (\vec{x}),
\ee
where
\be
h_ {\pm} = \frac{3}{2}\left( 1\pm \sqrt{1-\frac{4m^2}{9H^2}}\right).
\ee
The dominant solution for late times is the second one, so we can write:
\be\label{sigmah}
\phi(\vec{x}, \tau) \approx \tau^{h_ {-}} \sigma_{-} (\vec{x}).
\ee
As seen in \cite{Biagetti:2013qqa}, by imposing conformal invariance over $\sigma_{-} (\vec{x})$, and using (\ref{tl}) for any conformal transformation, say, for dilatations, we deduce that the field $\sigma$ has conformal weight
\be
\Delta_{\sigma} = h_ {-} = \frac{3}{2}\left( 1- \sqrt{1-\frac{4m^2}{9H^2}}\right).
\ee
In general, if slow roll inflation is assumed, it can be show that a scalar field perturbation propagating in a nearly de Sitter space has a conformal weight  of the form $\Delta_{\delta\phi}\sim \cal{O}(\epsilon,\eta)$, where $\epsilon$ and $\eta$ are the standard slow roll parameters (see {\it e.g.} \cite{Riotto:2002yw}).

Now, we begin to study the model given by the action (\ref{ccsv}). 
In the context that we frame our discussion, the scalar field could be either the inflaton or an auxiliary field present during inflation\footnote{In the particular case in which the scalar is an auxiliary field, other than the inflation, it need to be a light field, i.e. $m\ll H$.}.  The vector field is an auxiliary subdominant field which affects the primordial curvature perturbation and can leave an imprint on the inflationary evolution. The scalar affect the evolution of the vector field through the coupling functions $f_i(\phi)$. The equations of motion for the vector field derived from the action (\ref{ccsv}) are:
\be\label{eomg}
\nabla_{\mu}\left( f_{1} (\phi) F^{\mu \nu}  +  f_{2} (\phi) \tilde{F}^{\mu \nu}\right) = 0 
\ee
which is complemented with the (non dynamical) Bianchi identity $\nabla_{\mu}  \tilde{F}^{\mu \nu} = 0$.
We also assume that the inflationary dynamics homogenise the scalar perturbations,  and then, we can approximate the scalar field as a time dependent function, this is $\partial_{i}\phi=0$. Accordingly, this implies that on the solutions of the inflationary scalar field $f_{1}(\phi) = f_{1} (\phi (\tau))$ and $f_{2}(\phi) = f_{2} (\phi (\tau))$,  then $\partial_{i}f_{1} = \partial_{i}f_{2}=0$. Finally, we use the gauge symmetry of the action to impose the Coulomb gauge, this is, we set: $A_{0}=0$ and $\partial_i A_i=0$.  With this choice, the equation of motion (\ref{eomg}) reduces to:
\be\label{eomf1f2}
\left(\frac{\partial^{2}}{\partial \tau^{2}}    -  \nabla^{2}  + \frac{1}{f_{1}}\frac{\partial f_{1} }{\partial \tau} \frac{\partial  }{\partial \tau} +    \frac{1}{f_{1}}\frac{\partial f_{2} }{\partial \tau}  \nabla \times   \right) \vec{A}(\tau, \vec{x}) = 0.
\ee
So far, aside of the restriction imposed by  stability $f_1>0$, we don't have any further restriction over the form of the coupling function $f_2$. Now, we make a further assumption in order to respect conformal invariance on the asymptotic region. Let's assume that the coupling functions are homogeneous functions of time, this is $f_1({\lambda \tau}) = \lambda^n f_1(\tau)$, $f_2({\lambda \tau}) = \lambda^m f_2(\tau)$. Then, applying the dilatation $\tau' = \lambda \tau$ and $\vec{x}'= \lambda \vec{x}$  to (\ref{eomf1f2}) we obtain
\be \left(\frac{\partial^{2}}{\partial \tau^{2}}    - \nabla^{2}  + \frac{1}{f_{1}(\tau)}\frac{\partial f_{1} (\tau)}{\partial \tau} \frac{\partial  }{\partial \tau} +    \lambda^{m-n}\frac{1}{f_{1}(\tau)}\frac{\partial f_{2}(\tau) }{\partial \tau}  \nabla \times \right)\vec{A}(\tau, \vec{x}) = 0. 
\ee 
In this way we see that a necessary condition for getting scale invariance is that the coupling functions are homogeneous and of the same order $n=m$ \cite{Valenzuela2016}. This is achieved if the couplings are power law functions proportional to each other. In this case, it is  possible to see that symmetry under inversions $x^{\mu}\rightarrow x^{\mu}/\vec{x}^2 $, which imply invariance under special conformal transformations, is also guaranteed, see Appendix \ref{A1}. Then, renaming $f_1=f$ and $n=-2\alpha$ because, as we say before, $f$ must be positive due to hamiltonian stability,  we get: 
\be\label{fintime}
\boxed{ f_2(\tau) = \gamma f(\tau) \quad \mbox{with} \quad f(\tau) \propto (-H \tau)^{-2\alpha},}
\ee  
where we have restored the Hubble constant for dimensional analysis.  In order to avoid strong coupling at super horizon evolution, in the following we assume that $\alpha<0$.
Now, defining the normalized canonical field $w_i$ as $w_i \equiv \sqrt{f} A_i $, the equation of motion (\ref{eomf1f2}) becomes
\be \label{eomxalpha}
\left(\frac{\partial^{2}}{\partial \tau^{2}}    -  \nabla^{2}  -  \frac{\alpha (\alpha + 1)}{\tau^2} - \frac{2\alpha \gamma}{\tau}    \nabla \times   \right) \vec{w}(\tau, \vec{x}) = 0.
\ee
Going to momentum space, choosing propagation along the $x$-axis $\vec{k} = (k, 0,0)$ and defining transverse polarizations as $w_{\pm} = {(w_y \pm i w_z)}/{\sqrt{2}}\,$, we obtain
\be \label{eomkalpha}
\left(\frac{\partial^{2}}{\partial \tau^{2}}    +  k^{2}  -  \frac{\alpha (\alpha + 1)}{\tau^2} + \sigma \frac{2\xi k}{\tau}      \right) {\tilde{w}_{\sigma}}(\tau, \vec{k}) = 0,
\ee 
where $\sigma = \pm 1$ denote the helicity of the mode. Then, the system is quantized by expanding in terms of the mode solutions $ {\tilde{w}_{\sigma}}$ as
\be
\vec{w} (\tau, \vec{x}) = \sum_{ \sigma = \pm} \int \frac{d^3k}{(2\pi)^{3/2}} \left[ \epsilon^{i}_{\sigma}(\vec{k}) \tilde{w}_{\sigma} (\tau, \vec{k}) \hat{a}_{\sigma} (\vec{k}) e^{i\vec{k}\cdot \vec{x}} + \mbox{c.c} \right]
\ee 
where the transverse polarization vectors are defined in such a way that:
\be \label{pol}
\vec{\epsilon}_{\sigma}(\vec{k}) \cdot \vec{k} = 0, \quad   \vec{k} \times \vec{\epsilon}_{\sigma}  = - i \sigma |\vec{k}| \vec{\epsilon}_{\sigma}, \quad \vec{\epsilon}_{\sigma}\cdot \vec{\epsilon}_{\sigma'}  =\delta_{\sigma, -\sigma'},  \quad {\vec{\epsilon}}^{\; *}_{\sigma}(\hat{k})= \vec{\epsilon}_{-\sigma} (\hat{k}) = \vec{\epsilon}_{\sigma} (-\hat{k}) .
\ee
Following the notation in \cite{Caprini:2014mja,Anber:2009ua}, $\xi = -\alpha \gamma$  is the parameter which determines the relative strength of the parity violation effects and, without loosing generality we choose $\gamma$ to be  positive.\footnote{ The choice of the sign of $\gamma$ determines which one of the two helicities is the growing one. If  $\gamma>0$ the increasing mode is denoted with a subscript ``$+$'' and, if $\gamma<0$ the growing will be denoted with a ``$-$''. %{With this choice the ``$+$'' helicity is %exponentially 
%amplified by a factor $e^{\pi\xi/2}$ \cite{Caprini:2014mja,Anber:2009ua}.}
} 
Reference \citep{Caprini:2014mja} study in detail the features of this model and describe analytically its asymptotic behavior, here, we recall some basics about the solutions of this system.  
The equation (\ref{eomkalpha}) can be solved analytically in terms of irregular $G$ and irregular $F$  Coulomb functions:
\be 
\tilde{w}_{{\sigma}} (\tau, \vec{k}) = C_1^{\sigma}(k) G_{-\alpha-1} (\sigma \xi, -k\tau)  + C_2^{\sigma}(k) F_{-\alpha-1}(\sigma \xi, -k\tau).
\ee
(Alternatively, we can use Whittaker functions $ M_{\alpha + 1/2}(-i {\sigma} \xi, 2i k\tau), W_{\alpha + 1/2}(-i {\sigma} \xi, 2i k\tau) $ functions to solve (\ref{eomkalpha}) with the imaginary argument $z= 2i k \tau$ but the limiting form of Coulomb functions is better described in tables of mathematical functions, and is directly expressed in the real argument $x= -k \tau$. See for instance chapter 14 of \cite{abramowitzstegun}  and section 33 of \cite{NIST:DLMF}.) The asymptotic behavior of the Coulomb function  can be found in  expressions (14.1.7) and (14.6.7) of \cite{abramowitzstegun} and (33.10.3) and (33.10.7) of \cite{NIST:DLMF}. Using those expressions we find that for scales with $\xi \gg |k\tau|$ and $\xi\gg 1$\footnote{This is the regime in which the term $\frac{2\xi k}{\tau}$ dominates over the term $k^2$ in Eq. (\ref{eomkalpha}). We neglect $k^2$ but we keep $\frac{2\xi k}{\tau}$ since $\xi \gg 1$.},  the  solution of (\ref{eomkalpha}) with positive helicity $\sigma = +1$,  is approached by Bessel $K$ and Bessel $I$ functions  
%in  in which $|k\tau| \ll \xi$ the $W$ function dominates and we will consider only this contribution. Now, if we take $|k\tau| \ll \xi$ in (\ref{eomkalpha}) we have solutions in terms of Bessel $\cal{I}$ functions:
\ba\nonumber 
\tilde{w}_{ {+}} (\tau, \vec{k}) &\approx& \sqrt{-2 \xi k \tau} \left[ C_{1}^{+}(k)  \frac{ 2 ( \xi)^{-\alpha -1}  }{ |\Gamma(-\alpha + i \xi)| } e^{ \frac{\pi  \xi}{2} } { K }_{-(1+2\alpha)}\left( \sqrt{-8  \xi k \tau} \right) \right. \\ \nonumber
&& \left.\quad + C_{2}^{+} (k)  \frac{  |\Gamma(-\alpha + i \xi)|  }{  2( \xi)^{-\alpha }   } e^{- \frac{\pi  \xi}{2} }  { I }_{-(1+2\alpha)}\left( \sqrt{-8  \xi k \tau} \right) \right] \\ \label{spxikt}
&& \approx  \frac{ 2 ( \xi)^{-\alpha -1}  C_{1}^{+}(k)  }{ |\Gamma(-\alpha + i \xi)| } \sqrt{-2 \xi k \tau}  e^{ \frac{\pi  \xi}{2} } { K }_{-(1+2\alpha)}\left( \sqrt{-8  \xi k \tau} \right),
\ea
whereas the solution with negative helicity $\sigma =-1$ is approached by Bessel $Y$ and Bessel $J$ functions
\ba\nonumber 
\tilde{w}_{ {-}} (\tau, \vec{k}) &\approx& \sqrt{-2 \xi k \tau} \left[  C_{1}^{-}(k)  \frac{ (-\pi) ( \xi)^{-\alpha -1}  }{  |\Gamma(-\alpha - i \xi)| } e^{- \frac{\pi  \xi}{2} } { Y }_{-(1+2\alpha)}\left( \sqrt{-8 \xi k \tau} \right) \right. \\ \nonumber
&& \left.\quad +  C_{2}^{-} (k)  \frac{   |\Gamma(-\alpha - i \xi)|  }{  2( \xi)^{-\alpha }   } e^{ \frac{\pi  \xi}{2} }  { J}_{-(1+2\alpha)}\left( \sqrt{-8 \xi k \tau} \right) \right] \\ \label{smxikt}
&& \approx  \frac{ (-\pi) ( \xi)^{-\alpha -1}  C_{1}^{-}(k)  }{  |\Gamma(-\alpha - i \xi)| }  \sqrt{-2 \xi k \tau} e^{- \frac{\pi  \xi}{2} }{ Y }_{-(1+2\alpha)}\left( \sqrt{-8 \xi k \tau} \right) .
\ea
The expressions above show the different behavior of the solutions of (\ref{eomkalpha}) for the different helicities $\sigma$. According with (\ref{spxikt}) and (\ref{smxikt}), the $\sigma = +1$ mode is enhanced by the exponential factor $ e^{ \frac{\pi  \xi}{2} }$  \cite{Caprini:2014mja,Anber:2009ua}, while the $\sigma = -1$ mode is suppressed by the factor  $ e^{- \frac{\pi  \xi}{2} }$. This different behavior of the modes is a feature linked to the intrinsic parity violating nature of the system due to the presence of the term $f_2 F \tilde{F}$  in (\ref{ccsv}), in other words, the different behavior of the mode solutions reflects the presence of the term $f_2 F \tilde{F}$. 
Moreover, using the asymptotic form of Bessel functions, it is straightforward to see that going to super horizon scales, when $|8 \xi k \tau| \ll 1$, the dominant terms of the solution goes like
%\be
%\tilde{w}_{{\sigma}} (\tau, \vec{k}) \approx \left[ C_1^{\sigma} (k)(-2  \xi k \tau)^{\alpha + 1}\Gamma(-1- 2\alpha )+  C_2^{\sigma} (k)(-2 \xi k \tau)^{-\alpha}\Gamma(1+ 2\alpha ) \right],
%\ee 
%which can be written as
\be\label{apmasymp}
\tilde{w}_{\sigma} (\tau, \vec{k}) \approx  \tilde{u}_{\sigma} (k)(-\xi H \tau)^{\alpha + 1} +  \tilde{v}_{\sigma} (k)(-\xi H \tau)^{-\alpha} ,
\ee
where we have restored $H$ and absorbed further dependence of $k, \xi, \alpha$ in the $\tilde{u}$ and $\tilde{v}$ functions. At this point we go back to coordinates space obtaining the asymptotic form:
\be\label{apxmasymp}
{w}_{\sigma} (\tau, \vec{x}) \approx  {u}_{\sigma} (\vec{x})(-\xi H \tau)^{\alpha + 1} +  {v}_{\sigma} (\vec{x})(-\xi H \tau)^{-\alpha}.
\ee
We have separated the time and space coordinates for super horizon scales, and now, we can define a conformal boundary field in this case. We can see that the dominant term depends on the value of the exponent $\alpha$. If $\alpha<-1/2$ the term $v_{\sigma}$ dominates, while, if  $\alpha>-1/2$ the term $u_{\sigma}$ is the one that dominates. If $\alpha>-1/2$ we demand that\\
\be \label{bfv}
\lim_{|\tau| \rightarrow 0} {w_{\sigma}} (\tau, \vec{x} ) = \tau^{-\alpha}v_{\sigma}(\vec{x}),
\ee 
 while, if $\alpha<-1/2$ we take the boundary condition
\be \label{bfu}
\lim_{|\tau| \rightarrow 0} {w_{\sigma}} (\tau, \vec{x} ) = \tau^{\alpha + 1}u_{\sigma}(\vec{x}).
\ee   
Now, we can calculate the conformal weight of those boundary fields by consider the dilatation transformation, the inversion is entirely analogous \cite{Valenzuela2016}. We see that the four dimensional canonical field $w_{\sigma}$ transforms under four dimensional dilatations $\tau' = \lambda \tau$, $x' = \lambda x$, $x'_{\sigma} = \lambda x_{\sigma}$ as a four dimensional vector:
\be
w'_{\sigma}=\frac{\partial x_{\sigma}}{\partial x'_{\sigma}} w_{\sigma}=\lambda^{-\Delta_w}w_{\sigma}=\lambda^{-1}w_{\sigma}\;.
\ee
Applying dilatation over (\ref{bfv}) we have:
\be
w'_{\sigma}=(\tau')^{-\alpha}v'_{\sigma}(\vec{x}')= \lambda^{-\alpha}(\tau)^{-\alpha}v'_{\sigma}(\vec{x}')=\lambda^{-1}w_{\sigma}=\lambda^{-1}(\tau)^{-\alpha}v_{\sigma}(\vec{x})\;,
\ee
which implies
\be\label{udil}
v'_{\sigma}(\vec{x}')= \lambda^{\alpha-1}v_{\sigma}(\vec{x})\;,
\ee
and comparing with (\ref{tl}) with $r=1$ and $d=3$ we conclude that the three dimensional boundary field $v_{\sigma}$ behave as a conformal field of weight:
\be\label{cwv}
\Delta_{v} = 1-\alpha.
\ee
Following the same steps we find that the boundary field (\ref{bfu}) transforms as a conformal field of weight
\be\label{cwu}
\Delta_{u} = \alpha+2.
\ee
To summarise, we have obtained that the interacting system (\ref{ccsv}) admit a conformal field theory representation at super horizon scales through the boundary vector fields $u_{\sigma}, v_{\sigma}$. 
Those fields are the same that the ones found in \cite{Biagetti:2013qqa} for the parity conserving case. This fact can be better understood if we take the strong limit $\xi k \tau\rightarrow 0$. In this limit, the $\xi$ parameter (introduced by the $f_2$ term) enter as a multiplicative factor and can be absorbed as a redefinition of the fields. 
%A remarkable fact of this results is that they coincide with the case $ f(\phi) F^{\mu \nu}F_{\mu \nu}$ studied in \cite{Biagetti:2013qqa}. 
Then, the results here, extend the asymptotic symmetries analysis from \cite{Biagetti:2013qqa}  to the model (\ref{ccsv}) with  the parity violating term $f_2= \gamma f$, this is
\ba \label{betal}
S_{\rm{\phi A}} = -\frac{1}{4}\int d^{4}x \sqrt{-g} f(\phi)  \left[  F^{\mu \nu}F_{\mu \nu} + \gamma \tilde{F}^{\mu \nu}F_{\mu \nu} \right]\;.
\ea
This model was studied in detail in \cite{Caprini:2014mja} (and more recently in \cite{Caprini:2017vnn}) in the context of inflationary magnetogenesis and also in \cite{Dimopoulos:2012av} where there is a comprehensive discussion of particle creation with this coupling. Naively, we could think that both cases, the parity conserving and the parity violating cases have the same conformal boundary field associated and that the statistical features of both theories, given by the correlation functions, are the same. Nevertheless, there is a crucial difference implied by the fact that the super horizon boundary symmetry group allow the spacial reflection $\vec{x}' = - \vec{x}$ as an element of the symmetry group. This fact is also reflected in the super horizon evolution of the solutions to the equations of motion as can be seen from the equation (\ref{eomkalpha}). In next section we will discuss  the structure of the correlation functions. 
%%%%%%%%%%%%%%%%%%%%%%%%%%%%
\subsection{Perturbative expansion}\label{PE}
%%%%%%%%%%%%%%%%%%%%%%%%%%%%
Now, we discuss the structure of the interaction terms (\ref{ccsv}) which is necessary to develop a perturbative analysis of the theory. As we are considering that the scalar field is the inflaton field, we have that in the spatially flat gauge, the variation of the scalar field is proportional to the primordial curvature perturbation: $\delta \phi = -\frac{\phi'}{{\cal H}} \zeta$. The variation of the vector part of  (\ref{betal}) give rise to terms of the form $\langle F_{\mu \nu}\rangle + \delta F_{\mu \nu} $. It is convenient to study this system in terms of the ``electric" and ``magnetic" components of the vector field defined as:
\be 
E_i = -\frac{\sqrt{f}}{a^2} {\partial_{\tau} A_i}, \quad \mbox{and} \quad B_i = \frac{\sqrt{f}}{a^2}  \epsilon_{ijk}\partial_j A_k,
\ee  
where we normalized the fields absorbing the kinetic function $f$.  {Here, it is important to mention that the presence of a non-zero {\it v.e.v} for a vector field,  introduce an explicit violation of isotropy, which, in principle,  is inconsistent with the isotropic de Sitter spacetime. This inconsistency can be solved,  if, for example, we work in a Bianchi background but then the formalism presented here does not work. This issue has been addressed at the literature \cite{Dulaney10,Gumrukcuoglu10,Watanabe10}, and it was shown that the anisotropy in the background coming from a  vector field is negligible compared with the statistical anisotropies in the correlation functions. For instance, for the two point correlation function,  the amount of statistical anisotropy $g_\zeta$, due to $E_i^{(0)}$,  is order $g_\zeta\sim {\cal O} (10^{-1})$ where $g_{\zeta}$ is the appropriate parameter to quantify departures from statistical isotropy for a quadrupole form of the power spectrum $P_{\zeta} =  P_{\zeta}^0 (1 + g_{\zeta} (\hat{n}\cdot\hat{k})^2)$, whereas,  the  background anisotropy, which is  characterized by the anisotropy parameter $\Sigma/H$, is order $\Sigma/H\sim  {\cal O} (10^{-7})$ \cite{Dulaney10,Gumrukcuoglu10,Watanabe10}. %We can see that the anisotropy from the background is negligible compared with the one coming from the vector field. 
This results allows us to study the anisotropic signs on the correlators sourced  by  $E_i^{(0)}$ which can lead to observable signs, in a safe way, in an isotropic de Sitter background since departures from background isotropy are negligible}. 

In terms of the electric and magnetic components, the Lagrangian of the vector part is
\be
{\cal L}_{\phi A} = -\frac{1}{2}(\vec{B}^2 - \vec{E}^2) +\gamma  \vec{E}\cdot  \vec{B}.
\ee
Expanding, we get the interaction Lagrangian
\ba
\delta{\cal L}_{\phi A} = a^4 \frac{f'(\phi_0)}{f(\phi_0)} \frac{\phi'}{{\cal H}} \zeta \left[-\vec{E}^{0} \cdot \delta \vec{ E }  - \frac{1}{2} \delta \vec{ E } \cdot \delta \vec{ E }  +  \frac{1}{2}\delta \vec{ B } \cdot \delta \vec{ B } - \gamma ( \vec{E}^{0}\cdot  \delta \vec{B} +  \delta \vec{B} \cdot  \delta \vec{E}  )\right],
\ea
where we used that the zero mode (the homogeneous part) of the magnetic field is null. The relative size of the vector perturbations depends of the post horizon evolution. As we saw in equation (\ref{apxmasymp}), the super horizon evolution depends of the exponent $\alpha$. For instance for $\alpha<-1/2$ the electric component is dominant and the magnetic field can be safely neglected. On the other hand, for $\alpha>-1/2$ the magnetic component is dominant and the electric field is neglected at super horizon evolution. Then, we relate the electric component with $u_{\sigma}$ and the magnetic component with the $v_{\sigma}$ boundary field.\\  
The interaction term $f(\phi)(F^2 + \gamma \tilde{F}F) $ has been discussed in detail in Ref. \cite{Bartolo:2015dga}. As we said before, considering the electric component of the vector field, the leading order terms of the interaction hamiltonian coincide with the one of the $f F^2$ model, this is: 
\ba\label{H1}
{\cal H}_1 &=& 4 a^4 E^{(\rm{0})}_i \int d^3k \left[ \delta E_i (\vec{k}, \tau ) +\gamma \delta B_i (\vec{k}, \tau ) \right] \zeta_{-\vec{k}}, \\ \label{H2}
{\cal H}_2 &=& 2 a^4  \int d^3k d^3p \left[ \delta E_i (\vec{k}) \delta E_i (\vec{p}) -  \delta B_i (\vec{k}) \delta B_i (\vec{p}) +2\gamma \delta B_i (\vec{k}) \delta E_i (\vec{p})  \right] \zeta_{-(\vec{k}+ \vec{p} )}. 
\ea
This interaction hamiltonian is the key element for the calculation of the evolution of the correlations of the field on this model. Via the ``in-in" formalism we can construct the time evolution of the correlations with the hamiltonians ${\cal H}_1$ and ${\cal H}_2$.  Nevertheless, we are going to follow a different approach in this paper, since we are mainly interested in the super horizon behavior of the correlations, then, we only invoke the asymptotic conformal symmetry group at super horizon scales to derive the structure of the correlations. For our purposes here, it is important to notice that the structure of the vertices derived from the interaction hamiltonian determine the form of the correlators, all the correlators are written as combinations and contractions of the terms present in (\ref{H1})-(\ref{H2}) and the form of this combinations is constrained and shaped by conformal symmetries.  \\
Models including a coupling of the type $f(\phi)\tilde{F}F$ has been studied recently with some interest. In particular, the pseudoscalar coupling $\phi  \tilde{F}F$ has very similar features to the ones discussed in this paper, it is a model that breaks invariance under rotation and parity transformations \cite{Shiraishi:2013kxa, Bartolo:2014hwa}.   Its relevance, in the context of inflationary physics, has been a matter of interest and debate, in particular, this model  offer a mechanism to enhance the production of  gravitational waves during the inflationary period (see for instance Ref. \cite{Namba:2015gja}). Some criticism on this model have raised due to the validity of its perturbative expansion and about the consistency of this mechanism to produce sizable primordial gravitational waves without  contradict current observational limits \cite{Ferreira:2014zia, Ferreira:2015omg}.  Nevertheless, by using the perturbative approach, it was shown that the model have an available parameter window, that is in agreement with the current observational constraints \cite{Namba:2015gja,Peloso:2016gqs}.
%%%%%%%%%%%%%%%%%%%%%%%%%%%%
%%%%%%%%%%%%%%%%%%%%%%%%%%%%
\section{The structure of the correlation functions}\label{CorF}
%%%%%%%%%%%%%%%%%%%%%%%%%%%%
%%%%%%%%%%%%%%%%%%%%%%%%%%%%
As we said, we are interested in the structure of the correlation functions for super horizon scales in the model (\ref{betal}). For this scales, we exploit the asymptotic conformal symmetry of de Sitter space to derive constraints over the form of the correlation functions at this regime. We follow the method presented in \cite{Biagetti:2013qqa} to calculate the correlations among scalars and vector fields in the $f(\phi) F^2$ using standard conformal field theory techniques and Ward identities for the conformal group. We extend this analysis to the model $f(\phi)(F^2 + \gamma \tilde{F}F)$ and discuss the effect of the inclusion of the parity violation term in the structure of the correlation functions.  

%%%%%%%%%%%%%%%%%%%%%%%%%%%%
\subsection{The Ward identities}
%%%%%%%%%%%%%%%%%%%%%%%%%%%%
Here, we briefly discuss the Ward identities. Ward identities are the principle that express the invariance under symmetry transformations, which, in our case, are transformations belonging to the conformal group. First of all, we implement  translational and rotational  invariance which are straightforward and their implication on the form of the correlators  is rather simple to understand.   
Due to translation invariance, the correlators respect momentum conservation, so, all them are written in momentum space like
\be
\langle \phi (\vec{k}_1)  \phi (\vec{k}_2) \cdots  \phi (\vec{k}_n)\rangle = F(\vec{k}_1, \vec{k}_2, \cdots \vec{k}_n)\delta( \vec{k}_{12\cdots n}).
\ee
where $\vec{k}_{12\cdots n} \equiv \vec{k}_{1} + \cdots +  \vec{k}_{n}$. Given that we include the pseudo-scalar coupling $f(\phi) F\tilde{F}$  we also allow for the inversion transformation $\vec{x}\rightarrow -\vec{x}$, then, we allow also for transformations with negative determinant $\rm{det} (R) =-1$, so, the correlators that involve vector fields, transform respecting the $O(3)$ symmetry group (in the vector indices).  For instance, if there are two vectors in the correlator, $V_i, V_j$, we have:
\be
\langle V_i (\vec{k}_1)  V_j (\vec{k}_2)  \rangle = G(k_1) \left( a(k_1)\delta_{ij} + b(k_1) \hat{k}_{1i} \hat{k}_{1j} + c(k_1) \epsilon_{ijl} \hat{k}_{1l} \right) \delta( \vec{k}_{12}),
\ee
and so on. The remaining part of the conformal group, are the dilatations and the SCT. They impose different, complementary restrictions on the form of the correlators. A beautiful and detailed presentation of the role of conformal transformations in the context of inflationary physics can be found in \cite{Maldacena:2011nz} and further discussion and details of the role of conformal invariance in momentum space in a more general context can be found in \cite{Bzowski:2013sza,Antoniadis:2011ib,Coriano:2013jba}. Here we restrict ourselves to recall the very basics of the subject. 
%%%%%%%%%%%%%%%%%%%%%%%%%%%%
\subsubsection{Dilatations}
%%%%%%%%%%%%%%%%%%%%%%%%%%%%
In the following, we are only interested in the correlations among scalar and vector perturbations but  the procedure for higher order rank tensor perturbations is similar. According with (\ref{tl}), the transformation rule for those perturbations are, respectively:
\ba\label{tlsv}
 \sigma' (\vec{x}') = \left| \frac{\partial x'^{l}}{\partial x^{m}}\right|^{-\frac{\Delta_{\sigma} }{d}} \sigma (\vec{x}') \quad \mbox{and} \quad
 v_i' (\vec{x}') =  \left| \frac{\partial x'^{l}}{\partial x^{m}}\right|^{\frac{1-\Delta_{ v} }{d}} \frac{\partial x^{j}}{\partial x'^{i}} v_j  (\vec{x}). 
\ea  
The Ward identity for dilatations is obtained by using the infinitesimal transformation 
\ba \label{infdil}
x'^i = x^i + \delta_{\lambda} x^i = x^i + \lambda x^i , 
\ea
%(\ref{infdil}) 
and the infinitesimal transformation for the fields
\be
\delta_{\lambda} v_i (\vec{x}) = v_i' (\vec{x}) -  v_i (\vec{x}) \simeq  - \lambda (\Delta_v + x^{k}\partial_{k}) v_i (\vec{x}). 
\ee
As an example, let's consider the two point correlation function for two vector perturbations. In momentum space, dilatation imply that the correlator transforms as
\ba
\delta_{\lambda} \langle v_{i} (\vec{k}_1) v_{j} (\vec{k}_2)\rangle'  & = & -\lambda \left[-3 + \sum_{a=1}^{2} \left( \Delta_{v}  -  \vec{k}_a \cdot   \frac{\partial}{\partial \vec{k}_a }  \right)  \right] \langle v_{i} (\vec{k}_1) v_{j} (\vec{k}_2) \rangle',
\ea
where the notation $\langle \rangle'$ means that the Dirac's $\delta$ function is factored out. Some details of this calculation can be found in appendix \ref{AA}. Then, invariance under dilatations implies the Ward identity for the two point correlator between vector perturbations: 
\ba
\left[-3 + \sum_{a=1}^{2} \left( \Delta_{v}  -  \vec{k}_a \cdot   \frac{\partial}{\partial \vec{k}_a }  \right)  \right] \langle v_{i} (\vec{k}_1) v_{j} (\vec{k}_2) \rangle' = 0.
\ea
In general for $N-$point correlators involving scalars and vectors we can write:
\ba\label{wdil}
\boxed{\left[-3(N-1) + \sum_{a=1}^{N} \left( \Delta_{a}  -  \vec{k}_a \cdot   \frac{\partial}{\partial \vec{k}_a }  \right)  \right] \langle \sigma(\vec{k}_1)\cdots  \sigma(\vec{k}_s) v_{i} (\vec{k}_{s+1})\cdots v_{j} (\vec{k}_N)\rangle' = 0.}
\ea 
where $\Delta_a$ is the conformal weight of the corresponding scalar or vector perturbation.
%%%%%%%%%%%%%%%%%%%%%%%%%%%%
\subsubsection{Special conformal transformation}
%%%%%%%%%%%%%%%%%%%%%%%%%%%%
Now, we study constraints over scalar-vector correlators in momentum space imposed by SCT. Further details of the derivation can be found in Appendix \ref{AA}. Infinitesimal SCT are written as 
\ba \label{infsct}
x'^i = x^i + \delta_{\vec{b}}x^i = x^i + b^i \vec{x}^2 - 2 x^i\vec{x}\cdot\vec{b},
\ea
which, generate the transformation of the conformal field 
\be
\delta_{\vec{b}} v_i (\vec{x})\simeq  \left[ \left(2\Delta_v (\vec{b}\cdot \vec{x}) + \vec{x}^2 (\vec{b}\cdot\vec{\partial}) -2(\vec{b}\cdot \vec{x})(\vec{x}\cdot\vec{\partial})\right) \delta^{j}_i -2 (x^{j} b_{i} -b^{j} x_{i}) \right] v_j (\vec{x}). 
\ee
Going to Fourier space, we can derive the transformation law for $N$-point correlation functions.This procedure involves several integration by parts and a careful manipulation of the terms involving the Dirac's delta function. Using the results of Appendix \ref{AA} we derive the Ward identity for SCT
\begin{subequations}
\begin{empheq}[box=\widefbox]{align}\label{wsct}\nonumber
\left[ \sum_{a=1}^{N} 2(\Delta_{a} -3) \partial_{k_a^i} + D^{i}_a  \right] \langle \sigma(\vec{k}_1)\cdots  \sigma(\vec{k}_s) v_{i_1} (\vec{k}_{s+1})\cdots v_{i_N} (\vec{k}_N)\rangle' &  \\      -2 \sum_{p=s+1}^{N}  \Sigma^{i j_p}{}_{i_p} \langle \sigma(\vec{k}_1)\cdots  \sigma(\vec{k}_s) v_{i_1} (\vec{k}_{s+1}) \cdots v_{j_p} (\vec{k}_{p}) \cdots v_{i_N} (\vec{k}_N)\rangle' & =0,
 \end{empheq}
\end{subequations}
where
\ba
 D^{i}_a =  k^{i}_{a} \partial^2_{k_a} - 2(\vec{k}_a\cdot \vec{\partial}_{k_a}) \partial_{k^{i}_a} \quad \mbox{and} \quad
 \Sigma^{i j_p}{}_{i_p} = \delta^{i j_p}\partial_{k^{i_p}_p} -   \delta^{i}_{i_p}\partial_{k^{j_p}_p}. 
\ea
It is worth to notice that in this case, the term on the second line of the Ward identity acts only over vector indices. In general, for higher rank tensor perturbations (higher spin fields), the transformation rule include this terms with the corresponding transformation matrix $\Sigma$. As an example, let's take a two point correlator involving two vector fields: 
\ba\label{wsct2}
\left[ \sum_{a=1}^{2} 2(\Delta_{a} -3) \partial_{k_a^i} + k^{i}_{a} \partial^2_{k_a} - 2(\vec{k}_a\cdot \vec{\partial}_{k_a}) \partial_{k^{i}_a}  \right] \langle  v_{i_1} (\vec{k}_{1}) v_{i_2} (\vec{k}_2)\rangle' &  \\  \nonumber 
 - 2   \Sigma^{i l}{}_{i_1} \langle  v_{l} (\vec{k}_{1}) v_{i_2} (\vec{k}_2)\rangle'  -  & 2 \Sigma^{i l}{}_{i_2} \langle  v_{i_1} (\vec{k}_{1}) v_{l} (\vec{k}_2)\rangle'   =0.
\ea
%%%%%%%%%%%%%%
\subsection{Two point function}
%%%%%%%%%%%%%%
Now we use the Ward identities to put constraints on the shape of the correlation functions. First, we consider the two point correlation function. In our case, we can consider three types of correlators: scalar-scalar, scalar-vector and vector-vector, but here we neglect the scalar-vector correlator, since we suppose that both fields are not correlated at the time of horizon crossing. The scalar-scalar correlator is
\be
\langle \delta \phi (\vec{k}_1)  \delta \phi (\vec{k}_2) \rangle = P_{\delta\phi} (\vec{k}_1)\delta( \vec{k}_{12}),
\ee
and after applying (\ref{wdil}) for two scalar perturbations, we have:
\[ \left[ -3 + 2\Delta_{\phi} - \vec{k}\cdot \frac{\partial}{\partial \vec{k}}\right] \langle \delta\phi(\vec{k}) \delta\phi(-\vec{k}) \rangle = 0. \]
Solving the above equation we get \cite{Antoniadis:1996dj,Strominger:2001pn,Valenzuela2016}:
\be
P_{\delta\phi} ({k})= A k^{-3 + 2\Delta_{\delta\phi}},
\ee
where $\Delta_{\delta\phi}$ is the conformal weight of the scalar field in the asymptotic super horizon region. We can see that this result is consistent with the scale dependence of the perturbations (\ref{sigmah}), but this time is consequence of assuming conformal invariance. No further restrictions over this function are found when we apply the invariance under SCT.   \\
For the vector perturbations, the power spectrum can be  written as
\be
\langle \delta A_i (\vec{k}_1)  \delta A_j (\vec{k}_2) \rangle =P_{ij} (\vec{k}) \delta( \vec{k}_{12}),
\ee
where $P_{ij}$ is invariant under the $O(3)$ group (in the vector indices). Then, taking into account the momentum conservation $\vec{k}_1 + \vec{k}_2 = 0$, the most general $O(3)$ invariant two-point function that we can have in momentum space is:   
\be
\langle \delta A_i (\vec{k})  \delta A_j (-\vec{k}) \rangle  \equiv P_{ij} (\vec{k}) = P ({k})(\delta_{ij} + b_1(k) \hat{k}_i \hat{k}_j +{b_2(k)\epsilon_{ija}\hat{k}_a }). 
\ee
where $b_1(k)$ and $b_2(k)$ are scale dependent functions to be determined. The power spectrum is real, so, taking into account that $A^{\dagger}_i (\vec{k}) = A_i (-\vec{k})$, by imposing the reality condition $ P_{ij} (\vec{k})^{\dagger} = P_{ji} (\vec{k})$:
\[ \langle A_i (\vec{k})  A_j (-\vec{k}) \rangle^{\dagger} =  \langle A^{\dagger}_i (\vec{k})  A^{\dagger}_j (-\vec{k}) \rangle =  \langle A_i (-\vec{k})  A_j (\vec{k}) \rangle = \langle A_j (\vec{k}) A_i (-\vec{k})   \rangle \]
we get  
\[ b_2(k) \epsilon_{jia}\hat{k}_a  = -b_2(k) \epsilon_{ija}\hat{k}_a= b^{\dagger}_2(k) \epsilon_{ija}\hat{k}_a \]
which implies that $b^{\dagger}_2(k)=-b_2(k) $, so, $b_2$ is an imaginary function which we can write as  $b_2(k) = i b(k)$ 
where $b(k)$ is a real function. On the other hand, by using gauge symmetry and choosing the gauge $A_0=0$ and $\nabla^i A_i=0$, which, in momentum space implies that $k^i A_i=0$, we get $b_1=-1$. With this, we can write the two-point vector correlator as:
\ba\label{psab}
P_{ab} (\vec{k}) \equiv \Pi_{ab} P(k) \equiv [\Delta_{ab} + i b(k) \epsilon_{abc}\hat{k}_{c}] P(k).
\ea
with
\be
\Delta_{ab} \equiv \delta_{ab} - \hat{k}_{a} \hat{k}_{b}.
\ee
Several properties and identities involving $\Pi_{ab}$ and $\Delta_{ab}$ can be found in appendix \ref{AB}. Those properties are useful to evaluate the Ward identities in momentum space. For scaling invariance, the equation (\ref{wdil}) reads:  
\be
 \Pi_{ij}  (-3+2\Delta_v - \vec{k}\cdot\partial_{\vec{k}} )P(k)- i\epsilon_{ijk}\hat{k}_k P(k) \vec{k}\cdot\partial_{\vec{k}}   b (k)= 0.
\ee
This leads to a system of two decoupled equations:
\be
(-3+2\Delta_v - \vec{k}\cdot\partial_{\vec{k}} )P(k)=0, \quad   \vec{k}\cdot\partial_{\vec{k}}   b(k) = 0,
\ee
with solutions:
\be
P(k)= \frac{A}{k^{3-2\Delta_v}}, \quad   b(k) =  \beta,
\ee
 With  $A$ and $\beta$ constants. With this result, we can rewrite the spectrum as:
 \be \label{pab}
 P_{ab} (\vec{k}) = A k^{p} [ \Delta_{ab} + i \beta \epsilon_{abc}\hat{k}_{c}], \quad \mbox{with}\quad p=2\Delta_v -3.
 \ee
It is possible verify that this solution is consistent with invariance under SCT. To do that, we need the first two derivatives of the spectrum, the details can be seen in the  appendix \ref{AB}. Using the results (\ref{1std}) to (\ref{2ndcnt2}) in this appendix,  we can get the necessary combination of derivatives in the Ward identity (\ref{wsct2}) for the SCT:
\ba\label{1part}
&& 2 (\Delta_v - 3) \partial_{i}P_{lm} + k^{i}\nabla^2 P_{lm} - 2\vec{k}\cdot \nabla  \partial_{i} P_{lm} \\ \nonumber
&=& A {k^{p-1}}  \left[-2  \hat{k}_{i} \Delta_{lm}  + (p+1)  (\Delta_{il} \hat{k}_{m} + \Delta_{im} \hat{k}_{l} ) +   4  \hat{k}_{i} \hat{k}_{l} \hat{k}_{m} \right]  \\  \nonumber 
&+&   i\beta  {k^{p-1}}   \epsilon_{mlc} \left[ (p+1) \delta_{ic} + (1-p) \hat{k}_{i}  \hat{k}_{c} \right].
\ea
where we have used $p = 2\Delta_v -3$ and we have separated the symmetric and antisymmetric parts. We notice that this term appears twice since the calculation of the Fourier transform give us two equal terms, one for each momentum $\vec{k}_1$ and $\vec{k}_2$. On the other hand, the combination of first derivatives in the second part of the Ward identity gives:
\ba
 &&2\left( \delta_{ni}\partial_{l} P_{nm} - \delta_{il}\partial_{n} P_{nm} + \delta_{ni} \partial_{m} P_{ln} - \delta_{im}\partial_{n} P_{ln} \right) \\ \nonumber
&=& 2A \frac{k^p}{k} \left[  -2 \hat{k}_{i} \Delta_{lm} + (p+1)  ( \hat{k}_{l} \Delta_{im} +  \hat{k}_{m} \Delta_{li} ) + 4 \hat{k}_{i} \hat{k}_{l} \hat{k}_{m} \right] \\ \nonumber
&+& i2\beta \frac{k^p}{k} \left[ \epsilon_{mci} (\delta_{lc} + (p-1) \hat{k}_{l}\hat{k}_{c})+ \epsilon_{cli} (\delta_{mc} + (p-1)  \hat{k}_{m}\hat{k}_{c}  ) \right]
\ea 
where we have used  (\ref{1std}), (\ref{1stcnt}) and we separate the symmetric and the antisymmetric parts.
Comparing with (\ref{1part}) we see that the symmetric parts cancels exactly. To see that the antisymmetric part cancels as well, we use the identity (\ref{ide}) on appendix \ref{AB2}. To summarize, by using the conformal symmetry for the two point correlator, we obtain (\ref{pab}) which gives us the structure and the scale dependence of the correlation function up to some constants, the amplitude $A$ and the factor $\beta$ in the antisymmetric part. \\
The factors $A$ and $\beta$ can't be determined by symmetry arguments but, we can say something about them by analysing the asymptotic behavior of the solutions. For instance, it has been shown, that the $\beta$ coefficient can be fixed by noticing that in the super horizon regime, the polarization $(+)$ is exponentially amplifying and therefore is the dominant contribution of the vector perturbations, which is a sign of parity violation \cite{Anber:2009ua,Caprini:2014mja}. Then,  the product $\langle \delta A_a (\vec{k})\delta A_b (-\vec{k})\rangle$ matches in the asymptotic limit with the contribution coming from the polarization $(+)$, this is:
\be
\langle \delta A_a (\vec{k})\delta A_b (-\vec{k})\rangle \approx | \delta A_+(\vec{k})|^2  {\epsilon^{*}}_a^{(+)}(\hat{k})  {\epsilon}_b^{(+)}(\hat{k}), 
\ee
with the polarization vectors as defined in section \ref{asymptotic_solutions}. Using (\ref{pol}), it is possible to see that the polarization vectors obey
\be
{\epsilon^{*}}_a^{(+)}(\hat{k})  {\epsilon}_b^{(+)}(\hat{k}) = \frac{1}{2}\left[ \Delta_{ab}  + i\epsilon_{abc} \hat{k}_c\right],
\ee
then, matching with the spectrum (\ref{pab}) we can fix $\beta=1$. It is worth to distinguish the two different asymptotic behaviours of the two types of vector field perturbations seen before. For electric fields and for magnetic fields we have respectively 
 \be 
 P^{v}_{ab} (\vec{k}) = A_{v} k^{2\alpha +1} [ \Delta_{ab} + i  \epsilon_{abc}\hat{k}_{c}], \quad \mbox{and }\quad P^{u}_{ab} (\vec{k}) = A_{u} k^{-(2 \alpha +1)} [ \Delta_{ab} + i  \epsilon_{abc}\hat{k}_{c}].
 \ee
To summarize, by imposing conformal symmetry, we have deduced the structure of the two point correlator, up to amplitude factors. In order to have the complete form of the correlator, we need the further information about the system, such as initial conditions or the asymptotic behavior of the fields. 
%%%%%%%%%%%%%%%%%%%%%%%%%%%%
\subsection{Three point function}
%%%%%%%%%%%%%%%%%%%%%%%%%%%%
Now, we consider a three point cross  scalar-vector-vector  correlator
\be
\langle \delta\phi (\vec{k}_1)  \delta A_i (\vec{k}_2) \delta A_j (\vec{k}_3)  \rangle = B^{\phi AA}_{ij}  (\vec{k}_1, \vec{k}_2, \vec{k}_3 ) \delta( \vec{k}_{123}).
\ee
This correlator is relevant one for tracking statistical anisotropies and parity violation signals, such as was shown in Ref.  \cite{Bartolo:2015dga}. In the same way as we did for the two point function, we can constrain this correlator by using the asymptotic conformal symmetries. First, the most general combination of vector with indices $i,j$ with $O(3)$ invariance is  
\ba
\langle \delta \phi (\vec{k}_1) \delta A_i (\vec{k}_2) \delta A_j (\vec{k}_3)  \rangle' &=& a_1 \delta_{ij} + a_2  \hat{k}_{2i}  \hat{k}_{3j} + a_3  \hat{k}_{3i}  \hat{k}_{2j} + a_4  \hat{k}_{2i}  \hat{k}_{2j} +a_5  \hat{k}_{3i}  \hat{k}_{3j} \\ \nonumber
&+& { b_1\epsilon_{ija}\hat{k}_{2a} + b_2 \epsilon_{ija}\hat{k}_{3a}  } + {  b_3   \hat{k}_{2i} \epsilon_{jab} \hat{k}_{2a} \hat{k}_{3b} + b_4  \hat{k}_{3i} \epsilon_{jab}\hat{k}_{2a} \hat{k}_{3b} } \\ \nonumber
&+& {  b_5  \hat{k}_{2j} \epsilon_{iab} \hat{k}_{2a} \hat{k}_{3b} + b_6  \hat{k}_{3j} \epsilon_{iab}\hat{k}_{2a} \hat{k}_{3b} }
\ea
where the $a_m$ and $b_m$ functions can depend of the magnitude of the vectors $k_m$ and the scalar products of those vectors. The correlator is separated in two independent parts, a symmetric part, composed by the $a_m$-terms and the antisymmetric part composed by the $b_m$-terms. Using the divergenless condition we get 2 equations: $k_{2i}\langle \delta \phi (\vec{k}_1)  \delta A_i (\vec{k}_2) \delta A_j (\vec{k}_3)  \rangle=0$  and $k_{3j}\langle \delta\phi (\vec{k}_1) \delta A_i (\vec{k}_2) \delta A_j (\vec{k}_3)  \rangle=0$ which gives us relations between the different terms. For the symmetric terms we get:  
\begin{align}
a_3 =\frac{a_2}{(\hat{k}_2 \cdot \hat{k}_3)^2}-\frac{a_1}{(\hat{k}_2 \cdot \hat{k}_3)}, \quad a_4= a_5 = -\frac{a_2}{(\hat{k}_2 \cdot \hat{k}_3)}
\end{align}
and 
\be
b_1=b_5(\hat{k}_2 \cdot \hat{k}_3)+b_6,  \quad \mbox{and} \quad b_2=b_3 +b_4(\hat{k}_2 \cdot \hat{k}_3)
\ee
for the antisymmetric ones. Additionally, the correlator is symmetric under  the simultaneous changes $i \leftrightarrow j$, $k_2 \leftrightarrow k_3$, which imply $b_5=-b_3$ and $b_6=-b_4$. We also have reality of the correlators: $\langle \delta\phi(k_1) \delta A_i(k_2) \delta A_j(k_3)\rangle^\dagger=\langle \delta \phi(k_1)  \delta A_j(k_3) \delta A_i(k_2) \rangle$ which tell us that $a_1$ and $a_2$ are real, while $b_3$ and $b_4$ are imaginary. Finally, we define the functions $I_1, I_2, I_3, I_4$ such that $a_1= I_1 + (\hat{k}_2 \cdot \hat{k}_3)I_2 $, $a_2 = (\hat{k}_2 \cdot \hat{k}_3)I_1$, $b_3 =I_3$ and $b_4 =  I_4$, so, we write the correlator as: 
\ba
\langle \delta \phi (\vec{k}_1)  \delta A_i (\vec{k}_2) \delta A_j (\vec{k}_3)  \rangle' &=& \sum_{m=1}^4 I_m T_{ij}^{(m)}, \quad \mbox{where}  \\
T_{ij}^{(1)}&=&  \delta_{ij}- {\hat{k}_{2i} \hat{k}_{2j}} - { \hat{k}_{3i}  \hat{k}_{3j}} + {(\hat{k}_2 \cdot \hat{k}_3)}   \hat{k}_{2i}  \hat{k}_{3j}, \\ 
T_{ij}^{(2)}&=&  \hat{k}_{3i}  \hat{k}_{2j} - (\hat{k}_2 \cdot \hat{k}_3)\delta_{ij},\\
T_{ij}^{(3)}&=& (  \hat{k}_{2a}- \hat{k}_{3a})\epsilon_{ija}+(\hat{k}_{3j}\epsilon_{iab} - \hat{k}_{2i}\epsilon_{jab} ) \hat{k}_{2a} \hat{k}_{3b},\\
T_{ij}^{(4)}&=&( \hat{k}_{3a}- \hat{k}_{2a})(\hat{k}_2 \cdot \hat{k}_3)\epsilon_{ija}+( \hat{k}_{3i}\epsilon_{jab}- \hat{k}_{2j}\epsilon_{iab}) \hat{k}_{2a} \hat{k}_{3b}.
\ea
Additionally, contracting  the identity  (\ref{ide}) with $ \hat{k}_{2}$ and $ \hat{k}_{3}$, subtracting and rearranging we realize that $T_{ij}^{(3)} =  T_{ij}^{(4)}$, 
so, we only need the three matrices  $T_{ij}^{(1)}$,  $T_{ij}^{(2)}$ and  $T_{ij}^{(3)}$ to determine the form of the bispectrum. The form factors $I_1, I_2, I_3$ depend of the length of the momenta $k_1, k_2, k_3$, or equivalently, of two momenta, say $k_2, k_3$ and the angle between them, this is $I_m = I_m(k_1, k_2, k_3)$. Due to momentum conservation, we express the $k_1$ dependence in terms of $k_2$ and $k_3$, with $k_1^2 = k_2^2 + k_3^2 + 2 \vec{k}_1\cdot\vec{k}_2$. Then, write the 3-point function as  
\ba\label{antz}
\langle \delta \phi (\vec{k}_1)  \delta A_i (\vec{k}_2) \delta A_j (\vec{k}_3)  \rangle'  &=& \sum_{m=1}^3I_m T_{ij}^{(m)} %= \sum_{p=1}^3\alpha_p (k_2 k_3)^{a_p}(\vec{k}_2 \cdot \vec{k}_3)^{b_p} T_{ij}^{(p)}
\ea
Applying the Ward identity for dilatations we get 
\be
\left[ -6 + 2 \Delta_{v} + \Delta_{\delta\phi}- ( \vec{k}_2 \cdot \partial_{\vec{k}_2} +  \vec{k}_3 \cdot \partial_{\vec{k}_3})\right] \sum_{m=1}^3I_m T_{ij}^{(m)}   = 0.
\ee
From this equations and using
\ba\nonumber
\vec{k}_a \cdot \partial_{\vec{k}_a}T_{ij}^{(m)} = 0, 
\ea
we see that the scalar form factors are homogeneous functions of the momenta $k_1, k_2, k_3$ with degree $D=2\Delta_v -6$, this is, the general solution for each scalar form factor is 
\be
I_m = \alpha_m (k_1)^{2\Delta_v +\Delta_{\delta\phi}-6}g \left(\frac{k_2}{k_1} ,\frac{k_3}{k_1} \right),
\ee
or, equivalently, by permuting the order of the momenta in this expression. The most general form of the function $g$ can be written as a series in the ratios $k_2/k_1$ and $k_3/k_1$
\be\label{Iseries}
I_m = \alpha_m (k_1)^{2\Delta_v +\Delta_{\delta\phi}-6} \sum_{p,q} a_{p,q} \left(\frac{k_2}{k_1}  \right)^p \left( \frac{k_3}{k_1} \right)^q  %=  \alpha_m (k_1)^{2\Delta_v -6-2p} \left( {k_2 k_3}  \right)^p .
\ee
with some constant coefficients $a_{p,q}$. The form of the series defining the function $g$ is constrained by invariance under SCT. A complete and detailed study of conformal constraints on the scalar form factors can be found in \cite{Antoniadis:2011ib,Coriano:2013jba} for scalar operators and in \cite{Bzowski:2013sza} for scalar, vector, and tensor operators. In these references it was found that the Ward identity associated with SCT is written as a generalized hypergeometric function of the two variables  $k_2/k_1$ and $k_3/k_1$ and the solutions for this equation are the Appell's functions $F_4$. The derivation of the general solution of the SCT Ward's identity is beyond the scope of this paper. Instead of that, we want to have an idea of the behavior of the super horizon limit of the solutions. To this end, we will consider the squeezed limit configuration $k_1\rightarrow 0$ and $\vec{k}_3 \rightarrow -\vec{k}_2$. In this limit, the mode $k_1$ is well deep in the super horizon regime, and it dominates the behavior of the correlator, so, we expect that our results work well in this limit. We then consider an approximate solution compatible with dilatations and SCT up to orders suppressed at super horizon regime. As an approximation, we consider a single term in the series expansion (\ref{Iseries})
\be
I_m = \alpha_m (k_1)^{2\Delta_v +\Delta_{\delta\phi}-6} \left(\frac{k_2}{k_1}  \right)^p \left( \frac{k_3}{k_1} \right)^q.
\ee
Applying the SCT to this scalar function
\be
\left[ \sum_{a=2}^{3} 2(\Delta_{a} -3) \partial_{k_a^i} + D^{i}_a  \right] I_m = 0
\ee
we get the solution $p=-3 + 2\Delta_v,\; q=0 $ (equivalently $q=-3 + 2\Delta_v, \; p=0$). Then, the scalar form factors obtained is:
\be
I_m =\frac{\alpha_m  }{k_1^{3-\Delta_{\delta\phi}} k_2^{3-2\Delta_v}}. %\quad  \mbox{or}  \quad I_m =\frac{\alpha_m  }{k_1^{3-\Delta_{\delta\phi} }k_3^{3-2\Delta_v}}. 
\ee 
In terms of the conformal weights (\ref{cwv}), (\ref{cwu}) of the magnetic and electric components respectively, we can recast our result as: 
\be
I_m^{(B)} =\frac{\alpha_m  }{k_1^{3-\Delta_{\delta\phi}} k_2^{1+2\alpha}} \quad  \mbox{and}  \quad I_m^{(E)} =\frac{\alpha_m  }{k_1^{3-\Delta_{\delta\phi}} k_2^{-1-2\alpha}}. 
\ee 
Putting everything together, the approximate solution for the cross correlator, which is compatible with super horizon conformal symmetry is written as
\ba\label{Isol}
\langle \delta \phi (\vec{k}_1)  \delta A_i (\vec{k}_2) \delta A_j (\vec{k}_3)  \rangle'  &=& \frac{ 1 }{k_1^{3-\Delta_{\delta\phi}} k_2^{3-2\Delta_v}} \left[ \alpha_1 \left( \delta_{ij}- {\hat{k}_{2i} \hat{k}_{2j}} - { \hat{k}_{3i}  \hat{k}_{3j}} + {(\hat{k}_2 \cdot \hat{k}_3)}   \hat{k}_{2i}  \hat{k}_{3j} \right) \right. \qquad \no\\
&+&  \alpha_2 \left( \hat{k}_{3i}  \hat{k}_{2j}  - (\hat{k}_2 \cdot \hat{k}_3)\delta_{ij}  \right) \no\\
&+& \left.  i \alpha_3 \left(  (  \hat{k}_{2a}- \hat{k}_{3a})\epsilon_{ija}+(\hat{k}_{3j}\epsilon_{iab} - \hat{k}_{2i}\epsilon_{jab} ) \hat{k}_{2a} \hat{k}_{3b} \right) \right].
\ea
Up to amplitude factors, this form share the same structure with the result found in \cite{Bartolo12} for $fF^2$  and \cite{Bartolo:2015dga} for $f(F^2 + \gamma F\tilde{F})$. It is also interesting to notice that (\ref{Isol}) is in agreement with previous results found in \cite{Jain:2012ga, Jain:2012vm} in the squeezed limit, i.e. when $k_1\rightarrow 0$. Additionally, in our case we have an extra parity violating  term $ I_3 T_{ij}^{(3)}$ which appears due to the presence of the axial coupling term $f F\tilde{F}$. By taking the limit $k_1\rightarrow 0$, $\vec{k}_3 \rightarrow -\vec{k}_2$ we get
\ba\label{Isolsqueezed}
\langle \delta \phi (\vec{k}_1)  \delta A_i (\vec{k}_2) \delta A_j (\vec{k}_3)  \rangle'  &=&\frac{ 1 }{k_1^{3-\Delta_{\delta\phi}} k_2^{3-2\Delta_v}}  \left[ (\alpha_1 + \alpha_2) \left( \delta_{ij}- {\hat{k}_{2i} \hat{k}_{2j}} \right)   +  2i \alpha_3  \hat{k}_{2a} \epsilon_{ija}\right]. \quad
\ea
By taking $\alpha_1=\alpha_2=\alpha_3$ we obtain
\ba\label{Isolsqueezed2}
\langle \delta \phi (\vec{k}_1)  \delta A_i (\vec{k}_2) \delta A_j (\vec{k}_3)  \rangle'  &=& \frac{ 2\alpha_1 }{k_1^{3-\Delta_{\delta\phi}} k_2^{3-2\Delta_v}} \left[ \left( \delta_{ij}- {\hat{k}_{2i} \hat{k}_{2j}} \right)   + i  \hat{k}_{2a} \epsilon_{ija}  \right].
\ea
which is written as 
\ba \label{Isolsqueezed3}
\boxed{ \langle \delta \phi (\vec{k}_1)  \delta A_i (\vec{k}_2) \delta A_j (\vec{k}_3)   \rangle'  \propto  k_{1}^{-\Delta_{\delta\phi}}P_{\delta \phi} (k_1) P_{ij} (\vec{k}_2). }
\ea
The form of this expression %in precise agreement 
agrees with the consistency relation derived in \cite{Jain:2012ga, Jain:2012vm}. The main difference is that here we have  an additional parity violating term coming from the coupling $f F\tilde{F} $. Nevertheless,  by using this method, and within the approximations employed, we were not able derive the suppression factor for the consistency relation, since, as we said before, the symmetries analysis give us the form of the correlators up to amplitude factors. In order to derive the complete expression of the correlators's limit, we would need the initial conditions and the precise asymptotic form of the fields which would imply to solve the complete dynamics of the system.  \\
The result for the squeezed limit can also be seen as a consequence of the operator product expansion of the conformal theory which tell us that the short distance limit structure of the correlators of a conformal field theory is completely determined  by the conformal weight of the fields. It was shown in \cite{Antoniadis:1996dj, Antoniadis:2011ib} that for a three point function of scalar operators, the squeezed limit behave as 
\be
\langle {\cal O}(\vec{k}_1) {\cal O}(\vec{k}_2) {\cal O}(\vec{k}_3) \rangle \propto k_1^{-3+\Delta_{\cal O}} \langle{\cal O}(\vec{k}_2) {\cal O}(-\vec{k}_2) \rangle. %\propto P_{\cal O} (k_1) P_{\cal O} (k_2) 
\ee
So, what we get in (\ref{Isolsqueezed3}) is an example of the operator product expansion including scalar and vector operators. As in the case of scalar operators, we get that the behaviour of the short distance limit is defined by the conformal weight of the fields. In our case, the conformal weight of the fields where derived by using the asymptotic dynamics of the fields with the coupling (\ref{betal}) and imposing conformal invariance in this region. Relying on the operator product expansion and on (\ref{Isolsqueezed3}) we can argue that for $N$-point correlators involving scalars and vectors coupled by the interaction term (\ref{betal}) we can derive the limit scaling expression 
\be
 \langle \sigma(\vec{k}_1)\cdots  \sigma(\vec{k}_s) v_{i_1} (\vec{k}_{s+1})\cdots v_{i_N} (\vec{k}_N)\rangle  \propto \frac{P_{\sigma} (k_1)}{ k_{1}^{\Delta_{\sigma}}}   \langle \sigma(\vec{k}_2)\cdots  \sigma(\vec{k}_s) v_{i_1} (\vec{k}_{s+1})\cdots v_{i_N} (\vec{k}_N)\rangle.
\ee

%%%%%%%%%%%%%%%%%%%%%%%%%%%%
%%%%%%%%%%%%%%%%%%%%%%%%%%%%
\section{Conclusions and final remarks}\label{Con}
%%%%%%%%%%%%%%%%%%%%%%%%%%%%
%%%%%%%%%%%%%%%%%%%%%%%%%%%%
In this paper we study the constraints imposed by conformal symmetry on the correlators of an interacting theory involving a scalar and a gauge invariant vector field during inflation with a coupling of the form $f_1(\phi)F^2 +f_2(\phi)\tilde{F}F$ which allows for parity violation.  To this end, we exploit the relation between the de Sitter symmetry group and the conformal group 
at super horizon scales. We follow the methods presented in \cite{Biagetti:2013qqa} which studied the  conformal symmetry constraints of the model with interaction term $f(\phi)F^2$ in an inflationary setup. The term $f(\phi)F^2$ provides a source of non diluting statistical anisotropies which is reflected in the structure of the correlation functions of the system. Realizing that the coupling $f(\phi)F^2$ respects conformal invariance at late times, Ref.  \cite{Biagetti:2013qqa} calculated the conformal weight of the fields and then obtained the form of the correlators by solving the Ward identities associated with conformal transformations.  
Following these ideas,  we extend this symmetry analysis to the model $f_1(\phi)F^2 +f_2(\phi)\tilde{F}F$.
As a result, we found that this model is compatible with exact conformal symmetry at the asymptotic future when the coupling functions are homogeneous functions of the conformal time with the same degree, this is, when $f_2 = \gamma f_1$ with constant $\gamma$. Then, the model 
\be\label{scm} f(\phi)(F^2 +\gamma \tilde{F}F) \ee
admits a conformal field theory representation at the asymptotic future. 
Remarkably, the asymptotic form of the solutions of (\ref{scm}) have the same conformal weight of the $f(\phi)F^2$ model and with them we can derive the form of the correlators of this theory. Interestingly enough, by using this procedure, we derived the correct form of the correlators when comparing with the recent literature on the subject, see for instance \cite{Bartolo12} for the $fF^2$ model and \cite{Bartolo:2015dga} for the $f(F^2+\gamma F\tilde{F})$ model. Moreover,  in the squeezed limit when all the modes are in the super horizon regime, and we were able to derive the relation (\ref{Isolsqueezed3}) for the cross correlation of scalar and vectors with a power law coupling to the conformal time. This result is related with the consistency relation obtained in \cite{Jain:2012ga, Jain:2012vm} for the cross correlator in the $f(\phi)F^2$ model. The main shortcoming of the method followed here is that we can't obtain the suppression factor appearing in the squeezed limit, since, as we said before, the symmetry analysis gives us the general structure of the correlators up to amplitude factors. If we want to obtain the correct amplitude factors in the correlators, we need also to solve the dynamics of the system with the appropriate initial conditions. 

The formalism used here allows us to describe general characteristics and features of the inflationary correlators relying only on the symmetries of the system as the fundamental guiding principle. As illustrative examples, we only considered here the two and three point correlators of scalar and vectors but the formalism is, of course, applicable to general higher rank perturbation correlators. It would be interesting to use this approach to derive the general form of scalar, vector and tensor correlators and to derive limits such as (\ref{Isolsqueezed3}) involving these perturbations. This could be relevant, for instance, to have a template, or a benching mark form for the correlators in the analysis of signatures of sourced primordial chyral gravitational waves and CMB polarization. We expect to further investigate these subjects elsewhere.

%%%%%%%%%%%%%%%%%%%%%%%
\section*{Acknowledgments}
This work has received funding from COLCIENCIAS grants numbers 123365843539 RC FP44842-081-2014 and 110671250405 RC FP44842-103-2016, from Universidad Antonio Nari\~no grant number 2017239 and from the European Union's Horizon 2020 research and innovation programme under the Marie Skłodowska-Curie grant agreements 674896 and 690575. JPBA thanks Universidad del Valle for its warm hospitality during several stages of this project.  
%%%%%%%%%%%%%%%%%%%%%%% 

%%%%%%%%%%%%%%%%%%%%%%%%%%%%-Appendix-%%%%%%%%%%%%%%%%%%%%%%%%%%%%

\appendix
\section{The action and the SCT}\label{A1}

In this appendix, we will show that the action (\ref{betal}) is invariant under SCT on super-Hubble scales. To do this we will take advantage of two the fact that the SCT is built as  a composition of three consecutive transformations: {\it inversion $\to$ translation $\to$ inversion}, they can be written as 
$$\frac{ {x'}^\mu}{x'^2} = \frac{x^\mu}{x^2} + b^\mu.$$ On the other hand, it is easy to see that the action is invariant under translations, so we only discuss the details of the  inversions \cite{Biagetti:2013qqa,Valenzuela2016}. By using the results of section \ref{asymptotic_solutions}, in particular the result after applying the dilatations, which is summarized in Eq. (\ref{fintime}), we can write the action, in terms of the field $w_i \equiv \sqrt{f} A_i $, as:
\begin{multline} \label{ccsvgca2}
S_{\rm{\phi A}}=-\frac{1}{2}\int  d^{3}x d\tau\left\{ -\dot{w}_{i}\dot{w}^i- \frac{\alpha(\alpha + 1)}{\tau^{2}}w_i w^i+\partial_i w_j(\partial^i w^j-\partial^j w^i) \right.\\
\left.-2 \gamma \varepsilon_{0ijk} \left( \frac{\alpha }{\tau } w^{i} + \dot{w}^{i}\right) \partial^{j} w^{k} \right\}.
\end{multline}
% where  integration by parts have being made. Next, the SCT can be constructed applying  inversions and translation \cite{Biagetti:2013qqa,Valenzuela2016} as
%\be
%\text{(Invertion)}\times \text{(Translation)}\times \text{(Invertion)}\nonumber\;,
%\ee
%so we use inversions in order to prove conformal invariance for the action. 
Now, in super-Hubble scales ($-\tau^2 \ll |\vec{x}|^2$), the inversions are written as:
\be \label{lt}
x'^\mu = \frac{x^\mu}{\left|\vec{x}\right|^2}\;,
\ee
with the Jacobian matrix and its inverse as follows 
\be
\frac{\partial x'^{\mu}}{\partial x^{\nu}} = \frac{ J_\mu^{\nu} }{\left|\vec{x}\right|^2} ,  \quad \frac{\partial x^{\mu}}{\partial x'^{\nu}} = \left|\vec{x}\right|^2 J_\nu^{\mu}, \quad \mbox{with}  \quad J_0^0=1, \quad \mbox{and} \quad J_i^j=\delta_i^j-\frac{2\delta_{ik}x^k x^j}{\left|\vec{x}\right|^2 }.
\ee
With this, we have the transformation law for a 3-dimensional vector
$$w'_i(x')=\left|\vec{x}\right|^{2}J_i ^j w_j(x),$$ 
and the transformation for the volume element $d^3x'd\tau'= \left(\left|\vec{x}\right|^2 \right)^{-4} d^3xd\tau$, the time and spatial derivatives:  $\partial'_0 = \left|\vec{x}\right|^2 \partial_0, \ \partial'_i = \left|\vec{x}\right|^2 J_i^j\partial_j$ and the Levi-Civita tensor
$$\varepsilon'_{0ijk}=\det\left(\left|\vec{x}\right|^2 J_\mu^\nu\right)\varepsilon_{0ijk},\label{epsi} \
 \ {\rm with} \ \det\left(J_\mu^\nu\right)=-1,$$
where we have used (\ref{tl}) and (\ref{lt}) and the properties of the Levi-Civita symbol. Putting the above expressions in (\ref{ccsvgca2}) and assuming that $\gamma$ is a constant\footnote{Actually, if $\gamma$ is not a constant, then the action is not invariant.}, we find that the action changes as
\ba
S'_{\rm{\phi A}}&=&-\frac{1}{2}\int %\frac{ 
d^{3}x' d\tau'
%}{(\left|\vec{x}\right|^2)^4}(\left|\vec{x}\right|^2)^{2+2\Delta_w}
\left\{ -\dot{w}'_{i}\dot{w}'^i- \frac{\alpha(\alpha + 1)}{\tau'^{2}}w'_i w'^i+\partial'_i w'_j(\partial'^i w'^j-\partial'^j w'^i)\right. \nonumber\\
&&-\left.2 \gamma \varepsilon'_{0ijk} \left( \frac{\alpha }{\tau' } w'^{i} + \dot{w}'^{i}\right) \partial'^{j} w'^{k}\right\}\no \\
&=&-\frac{1}{2}\int \frac{ 
d^{3}x d\tau}{(\left|\vec{x}\right|^2)^4}(\left|\vec{x}\right|^2)^{4}
\left\{ -\dot{w}_{i}\dot{w}^i- \frac{\alpha(\alpha + 1)}{\tau^{2}}w_i w^i+\partial_i w_j(\partial^i w^j-\partial^j w^i) \right.\no\\
&& \left.+2 \gamma \varepsilon_{0ijk} \left( \frac{\alpha }{\tau } w^{i} + \dot{w}^{i}\right) \partial^{j} w^{k} \right\}\no,
%\ = \ S_{\rm{\phi A}}.
%&-&\frac{4 \gamma}{\left|\vec{x}\right|^2}\left(\Delta_w-1\right)\left( \frac{\alpha }{\tau } w_{i} + \dot{w}_{i}\right)\varepsilon^{0ijk}x_j w_k \nonumber \\
%&+&\left.\frac{4\left(\Delta_w-1\right)^2}{(\left|\vec{x}\right|^2)^2}w_i w_j\left(\delta^{ij}\left|\vec{x}\right|^2-x^ix^j\right)%+4\frac{\left(\Delta_w-1\right)}{(\left|\vec{x}\right|^2)}w_k\partial_j w_k\left(x^j \delta^{ik}-x^k \delta^{ij}\right)\right\}\;, %\nonumber
\ea 
where we get a minus sign in the term proportional to $\gamma$. Now, we do a translation which leaves the action invariant and a second inversion which introduces a new minus sign in the last term, then, the action is left invariant under special conformal transformations. 

%%%%%%%%%%%%%%%%%%%%%%%%%%%%%%%
\section{Details of the calculation of the Ward identities}\label{AA}
%%%%%%%%%%%%%%%%%%%%%%%%%%%%%%%
Here we show the details of the derivation of the Ward identities for dilations and SCT. For all the calculations we apply the infinitesimal form of (\ref{tl}) to the scalar and vector fields. Then, we do a expansion to first order in powers of the corresponding symmetry generators and then we build the correlators and their  equations in coordinate space. Finally, we go to Fourier space and demand symmetry invariance to derive the corresponding  Ward identities.
%%%%%%%%%%%%%%%%%%%
\subsection{Dilatations}\label{AAa}
%%%%%%%%%%%%%%%%%%%
%In this section we will calculate the Ward identities related to the dilations for the two and three point correlation functions.
%%%%%%%%%%%%%%%%5
\subsubsection{Dilations 2-point function Ward identities}
%%%%%%%%%%%%%%%%%
 Applying an infinitesimal change of coordinates, such as $x'^i=x^i+\lambda  x^i $,  the conformal transformation (\ref{tl}) for a vector field up to first order in $\lambda$ is:
\be\label{gvtrans}
v_i' (\vec{x}') =  \left| \frac{\partial x'^{l}}{\partial x^{m}}\right|^{\frac{1-\Delta_{v} }{d}} \frac{\partial x^{j}}{\partial x'^{i}} v_j  (\vec{x}) \simeq (1 + (n-\Delta -1) \lambda)\delta^{j}_{i}\;.
\ee
Which allow us to write the variation of the field as:
\be
\delta_{\lambda} v_i (\vec{x}) = v_i' (\vec{x}) -  v_i (\vec{x}) \simeq  - \lambda (\Delta + x^{k}\partial_{k}) v_i (\vec{x}), 
\ee
and then the variation of the two point correlator: 
\be
\delta_{\lambda} \langle v_{i} (\vec{x}_1) v_{j} (\vec{x}_2)\rangle = - \lambda \sum_{a=1}^2  \left(\Delta + x_{a}^{l}\frac{\partial}{\partial {x_{a}^l}} \right)  \langle v_{i} (\vec{x}_1) v_{j} (\vec{x}_2)\rangle.
\ee
Going to Fourier space and after some algebra, the above expression reads:
\ba
&& \delta_{\lambda} \langle v_{i} (\vec{x}_1) v_{j} (\vec{x}_2)\rangle   =  \int \frac{d^3 k_1}{(2\pi)^3} \frac{d^3 k_2}{(2\pi)^3}  e^{i(\vec{x}_1\cdot \vec{k}_1 + \vec{x}_2\cdot \vec{k}_2)}  \delta_{\lambda} \langle v_{i} (\vec{k}_1) v_{j} (\vec{k}_2)\rangle \\ \nonumber
&& = -\lambda \int \frac{d^3 k_1}{(2\pi)^3} \frac{d^3 k_2}{(2\pi)^3}  e^{i(\vec{x}_1\cdot \vec{k}_1 + \vec{x}_2\cdot \vec{k}_2)}\delta (\vec{k}_1 + \vec{k}_2) \left[-3 + \sum_{a=1}^{2} \left( \Delta  -  \vec{k}_a \cdot   \frac{\partial}{\partial \vec{k}_a }  \right)  \right] \langle v_{i} (\vec{k}_1) v_{j} (\vec{k}_2) \rangle'  ,
\ea
which result in the next Ward identity:
\be
\boxed{
\delta_{\lambda} \langle v_{i} (\vec{k}) v_{j} (-\vec{k})\rangle'  = -\lambda \left[-3 + 2\Delta  -  \vec{k}\cdot \frac{\partial}{\partial \vec{k}}  \right] \langle v_{i} (\vec{k}) v_{j} (-\vec{k}) \rangle' \;.
}
\ee
%%%%%%%%%%%%%%%%%%%%%%%%%%%%
\subsubsection{Dilations 3-point function Ward identities}
%%%%%%%%%%%%%%%%%%%%%%%%%%%%%%
In this section we calculate the Ward identity for a 3-point correlator of the form $\langle \phi (\vec{x}_1) v_i (\vec{x}_2) v_j (\vec{x}_3)\rangle$. The infinitesimal conformal transformation for dilations for scalar and vector fields, this is  $\phi' (\vec{x}') \simeq \phi(\vec{x}) - \lambda (\Delta_\phi + x^{k}\partial_{k})\phi (\vec{x})\;$ and $v_i' (\vec{x}') \simeq v_i (\vec{x}) - \lambda (\Delta_v + x^{k}\partial_{k}) v_i (\vec{x})\;,$ allow us to write the variation of the three point function as:
\be
\label{delta3V}
\delta_\lambda \langle \phi' (\vec{x}'_1) v_i' (\vec{x}'_2) v_j' (\vec{x}'_3)\rangle=\\
 - \lambda (\Delta_\phi+2 \Delta_v + \sum_{l=1}^3x^{k}_l\frac{\partial}{\partial x_{l}^k})\langle \phi (\vec{x}_1)v_i (\vec{x}_2)v_j(\vec{x}_3)\rangle\;.
\ee
Now, doing the Fourier transforms of the fields $\phi(\vec{x})=\frac{1}{(2\pi)}\int d^3 k \tilde{\phi}(\vec{k}) e^{i \vec{k}\cdot\vec{x}}$ and $v_i(\vec{x})=\frac{1}{(2\pi)}\int d^3 k \tilde{v}_i(\vec{k}) e^{i \vec{k}\cdot\vec{x}} \label{vecftrans}$, the Fourier transformation of  (\ref{delta3V}) is written as:
\begin{multline}\label{delta3Vkk}
\frac{1}{(2\pi)^2}\int  d^3k_2 d^3 k_3 e^{i \vec{k}_2\cdot (\vec{x}_2-\vec{x}_1)+i\vec{k}_3\cdot (\vec{x}_3-\vec{x}_1)} \delta_\lambda \langle  \tilde{\phi} \tilde{v}_i  \tilde{v}_j \rangle'=\\
- \lambda \frac{1}{(2\pi)^2}\int d^3k_2 d^3k_3 e^{i \vec{k}_2\cdot (\vec{x}_2-\vec{x}_1)+i\vec{k}_3\cdot (\vec{x}_3-\vec{x}_1)}  \left(-6+\sum_{m=1}^3 \Delta_m + k_2^l\frac{\partial}{\partial k_2^l}+k_3^l \frac{\partial}{\partial k_3^l}\right)\langle \tilde{\phi}\tilde{v}_i  \tilde{v}_j\rangle' \;,
\end{multline}
which allow us to write the corresponding Ward identity for the 3 point function as
\be
\boxed{
\delta_\lambda \langle  \tilde{\phi}(\vec{k}_1) \tilde{v}_i  (\vec{k}_2) \tilde{v}_j (\vec{k}_3)\rangle'=\left(-6+\sum_{m=1}^3 \Delta_m + k_2^l\frac{\partial}{\partial k_2^l}+k_3^l \frac{\partial}{\partial k_3^l}\right)\langle  \tilde{\phi}(\vec{k}_1) \tilde{v}_i  (\vec{k}_2) \tilde{v}_j (\vec{k}_3)\rangle' \;.}
\ee
%\newpage
%%%%%%%%%%%%%%
\subsection{SCT}\label{AAb}
%%%%%%%%%%%%%%
%In this section we center our attention in the SCT.  
%%%%%%%%%%%%%%%%%%%%%%%%%
\subsubsection{SCT 2-point function}\label{scttp}
%%%%%%%%%%%%%%%%%%%%%%%%%%%%
By using the infinitesimal SCT, $\delta x^i=x^2b^i-2x^i (\vec{x}\cdot \vec{b})$, the transformation (\ref{tl}) for a vector $v_i(\vec{x})$ up to first order in $\vec{b}$ is
 \be\label{gvtrans}
v_i' (\vec{x}') = v_i(\vec{x})-2M_i^a(\vec{x}) v_a(\vec{x})+2 \Delta_v  (\vec{x}\cdot \vec{b}) v_i(\vec{x}')-\delta x^a \partial_a v_i (\vec{x}') \;,
\ee
where we have used the fact that, in this case, $\partial x'^{i}/\partial x^{j} \simeq \delta^{i}_{j}+2M^i_j-2\delta^i_j(\vec{x}\cdot \vec{b})$ %($\partial x^{i}/\partial x'^{j} \simeq\delta^{i}_{j}-2M^i_j+2\delta^i_j(\vec{x}\cdot \vec{b})$) 
and then
$\left|\partial x'^{i}/\partial x^{j}\right|^{\frac{r-\Delta_{n} }{d}} \simeq (1-6 (\vec{x}\cdot \vec{b}) )^{\frac{r-\Delta_n}{d}} \simeq 1 -6 \frac{(r-\Delta_n)}{d} (\vec{x}\cdot \vec{b})\;$ with $M^i_j=x_j b^i-x^i b_j$. With this, and using (\ref{gvtrans}), the two-point correlation function is:
  \begin{multline}\label{2gvtranscor}
\delta_b \langle v_i (\vec{x}_1)v_j (\vec{x}_2)\rangle =-2M_{i}^a(\vec{x}_1) \langle v_a(\vec{x}_1) v_j(\vec{x}_2)\rangle+2 \Delta_v  (\vec{x}_1\cdot \vec{b}) \langle v_i(\vec{x}_1)v_j(\vec{x}_2)\rangle-\delta x^a_1 \frac{\partial}{\partial x^a_1} \langle v_i(\vec{x}_1)v_j(\vec{x}_2)\rangle- \\
-2M_{j}^a(\vec{x}_2) \langle v_i (\vec{x}_1) v_a(\vec{x}_2)\rangle +2 \Delta_v  (\vec{x}_2\cdot \vec{b}) \langle v_i(\vec{x}_1)v_j(\vec{x}_2)\rangle-\delta x^a_2 \frac{\partial}{\partial x^a_2} \langle v_i(\vec{x}_1)v_j(\vec{x}_2) \rangle \;.
\end{multline} 
%\newpage
and in Fourier space, 
\begin{multline}\label{wardsctft}
\int d^3k_1 d^3 k_2 e^{i(\vec{k}_1\cdot \vec{x}_1+\vec{k}_2\cdot \vec{x}_2)}\delta_b \langle \tilde{v}_i (\vec{k}_1) \tilde{v}_j (\vec{k}_2)\rangle=\int d^3k_1 d^3 k_2 \left[ \right.\\
-2M_{i}^a(\vec{x}_1) \langle \tilde{v}_a(\vec{k}_1) \tilde{v}_j(\vec{k}_2)\rangle+2 \Delta_v  (\vec{x}_1\cdot \vec{b}) \langle \tilde{v}_i(\vec{k}_1)\tilde{v}_j(\vec{k}_2)\rangle-\delta x^a_1 \frac{\partial}{\partial x^a_1} \langle \tilde{v}_i(\vec{k}_1)\tilde{v}_j(\vec{k}_2)\rangle-\\
\left. -2M_{j}^a(\vec{x}_2) \langle \tilde{v}_i (\vec{k}_1) \vec{v}_a(\vec{k}_2)\rangle +2 \Delta_v  (\vec{x}_2\cdot \vec{b}) \langle \tilde{v}_i(\vec{k}_1)\tilde{v}_j(\vec{k}_2)\rangle-\delta x^a_2 \frac{\partial}{\partial x^a_2} \langle \tilde{v}_i(\vec{k}_1)\tilde{v}_j(\vec{k}_2) \rangle \right]e^{i(\vec{k}_1\cdot \vec{x}_1+\vec{k}_2\cdot \vec{x}_2)}\;.
\end{multline}
Now, we will use the fact that 
\be
\frac{\partial}{\partial k^l_p}e^{i(\vec{k}_1\cdot\vec{x}_1+\vec{k}_2\cdot\vec{x}_2)}=i x_{(p)l}e^{i(\vec{k}_1\cdot\vec{x}_1+\vec{k}_2\cdot\vec{x}_2)},
\quad \frac{\partial}{\partial x^l_p}e^{i(\vec{k}_1\cdot\vec{x}_1+\vec{k}_2\cdot\vec{x}_2)}=i k_{(p)l}e^{i(\vec{k}_1\cdot\vec{x}_1+\vec{k}_2\cdot\vec{x}_2)},
\ee
where $p=1,2$ and integration by parts in just one line in the RHS of (\ref{wardsctft}), knowing, that the other line can be computed by changing $1 \leftrightarrow 2$. Each one of the terms in the  first line of the RHS of (\ref{wardsctft}) are written as:
\bea
e^{i(\vec{k}_1\cdot\vec{x}_1+\vec{k}_2\cdot\vec{x}_2)} 2 \Delta_v  (\vec{x}_1\cdot \vec{b}) \langle \tilde{v}_i(\vec{k}_1)\tilde{v}_j(\vec{k}_2)\rangle
&=&2i e^{i(\vec{k}_1\cdot \vec{x}_1+\vec{k}_2\cdot \vec{x}_2)}    \Delta_v (b^l \frac{\partial}{\partial k^l_1}) \langle \tilde{v}_i(\vec{k}_1)\tilde{v}_j(\vec{k}_2)\rangle\;,\no\\
-2M_{i}^a(\vec{x}_1) \langle \tilde{v}_a(\vec{k}_1) \tilde{v}_j(\vec{k}_2)\rangle e^{i(\vec{k}_1\cdot\vec{x}_1+\vec{k}_2\cdot\vec{x}_2)}&=&-2i e^{i(\vec{k}_1\cdot\vec{x}_1+\vec{k}_2\cdot\vec{x}_2)} \tilde{M}^a_i  \langle \tilde{v}_a(\vec{k}_1) \tilde{v}_j(\vec{k}_2)\rangle \;,\no\\
-e^{i(\vec{k}_1\cdot\vec{x}_1+\vec{k}_2\cdot\vec{x}_2)}\delta x^a_1 \frac{\partial}{\partial x^a_1} \langle \tilde{v}_i(\vec{k}_1)\tilde{v}_j(\vec{k}_2)\rangle&=&i e^{i(\vec{k}_1\cdot\vec{x}_1+\vec{k}_2\cdot\vec{x}_2)}\left[(\vec{k}_1\cdot\vec{k_1}) \delta^{al} \frac{\partial}{\partial k_1^a}\frac{\partial}{\partial k_1^l}-2k_1^a \frac{\partial}{\partial k_1^a} b^l\frac{\partial}{\partial k_1^l}\right.\no\\
&&\left.-6b^a\frac{\partial}{\partial k_1^a}\right]\langle \tilde{v}_i(\vec{k}_1)\tilde{v}_j(\vec{k}_2)\rangle\;.
\eea
where $\tilde{M}^a_i=b^a \frac{\partial}{\partial k_1^i}-b_i \frac{\partial}{\partial k_{(1)a}}$. Follow the same procedure for the second line of the RHS of (\ref{wardsctft}) one found that this equation can be written as:
\begin{multline}\label{wardsctftol}
\int d^3k_1 d^3 k_2 e^{i(\vec{k}_1\cdot \vec{x}_1+\vec{k}_2\cdot \vec{x}_2)}\delta_b \langle \tilde{v}_i (\vec{k}_1) \tilde{v}_j (\vec{k}_2)\rangle=i \int d^3k_1 d^3 k_2  e^{i(\vec{k}_1\cdot \vec{x}_1+\vec{k}_2\cdot \vec{x}_2)}\\
 \left[ \right. 2(\Delta_v-3)  (b^l \frac{\partial}{\partial k^l_1})+(\vec{b}\cdot\vec{k_1}) \delta^{al} \frac{\partial}{\partial k_1^a}\frac{\partial}{\partial k_1^l}-2k_1^a \frac{\partial}{\partial k_1^a} b^l\frac{\partial}{\partial k_1^l}+\\
\left.+2(\Delta_v-3)  (b^l \frac{\partial}{\partial k^l_2})+(\vec{b}\cdot\vec{k_2}) \delta^{al} \frac{\partial}{\partial k_2^a}\frac{\partial}{\partial k_2^l}-2k_2^a \frac{\partial}{\partial k_2^a} b^l\frac{\partial}{\partial k_2^l} \right]\langle \tilde{v}_i(\vec{k}_1)\tilde{v}_j(\vec{k}_2)\rangle\\
-2\left[\left(b^a \frac{\partial}{\partial k_1^i}-b_i \frac{\partial}{\partial k_{(1)a}}\right) \langle \tilde{v}_a(\vec{k}_1) \tilde{v}_j(\vec{k}_2)\rangle+\left(b^a \frac{\partial}{\partial k_2^j}-b_j \frac{\partial}{\partial k_{(2)a}}\right)\langle \tilde{v}_i (\vec{k}_1) \tilde{v}_a(\vec{k}_2)\rangle\right]\;.
\end{multline}
Now, recalling that $\langle \tilde{v}_i (\vec{k}_1) \tilde{v}_a(\vec{k}_2)\rangle=\delta(\vec{k}_{12})\langle \tilde{v}_i (\vec{k}_1) \tilde{v}_a(\vec{k}_2)\rangle'$, and taking into account that the correlators are invariant under dilatations and are covariant under rotations, the operator in front of each $\langle \tilde{v}_i (\vec{k}_1) \tilde{v}_a(\vec{k}_2)\rangle$ does not affect the delta function. Then we have
\begin{align}\label{wardsctftodel}
\int d^3k_1 d^3 k_2 &\delta(\vec{k}_1+\vec{k}_2) e^{i(\vec{k}_1\cdot \vec{x}_1+\vec{k}_2\cdot \vec{x}_2)}\delta_b \langle \tilde{v}_i (\vec{k}_1) \tilde{v}_j (\vec{k}_2)\rangle'=i \int d^3k_1 d^3 k_2 \delta(\vec{k}_1+\vec{k}_2)  e^{i(\vec{k}_1\cdot \vec{x}_1+\vec{k}_2\cdot \vec{x}_2)}\nonumber\\
 &\left[ \right. 2(\Delta_v-3)  (b^l \frac{\partial}{\partial k^l_1})+(\vec{b}\cdot\vec{k_1}) \delta^{al} \frac{\partial}{\partial k_1^a}\frac{\partial}{\partial k_1^l}-2k_1^a \frac{\partial}{\partial k_1^a} b^l\frac{\partial}{\partial k_1^l}+\nonumber\\
&\left.+2(\Delta_v-3)  (b^l \frac{\partial}{\partial k^l_2})+(\vec{b}\cdot\vec{k_2}) \delta^{al} \frac{\partial}{\partial k_2^a}\frac{\partial}{\partial k_2^l}-2k_2^a \frac{\partial}{\partial k_2^a} b^l\frac{\partial}{\partial k_2^l} \right]\langle \tilde{v}_i(\vec{k}_1)\tilde{v}_j(\vec{k}_2)\rangle' \nonumber\\
&-2\left[\left(b^a \frac{\partial}{\partial k_1^i}-b_i \frac{\partial}{\partial k_{(1)a}}\right) \langle \tilde{v}_a(\vec{k}_1) \tilde{v}_j(\vec{k}_2)\rangle'+\left(b^a \frac{\partial}{\partial k_2^j}-b_j \frac{\partial}{\partial k_{(2)a}}\right)\langle \tilde{v}_i (\vec{k}_1) \tilde{v}_a(\vec{k}_2)\rangle'\right]\;.
\end{align}
Finally, evaluating the delta functions and renaming $\vec{k}_1=\vec{k}$, $\vec{k}=-\vec{k}_2$, we get
\begin{multline}
 \int d^3k e^{i\vec{k}\cdot(\vec{x}_1-\vec{x}_2)}\delta_b \langle \tilde{v}_i (\vec{k}) \tilde{v}_j (-\vec{k})\rangle'=\\
i \int d^3k  e^{i\vec{k}\cdot( \vec{x}_1- \vec{x}_2)}\left\{ \left[4 (\Delta_v-3)  (b^l \frac{\partial}{\partial k^l})+2(\vec{b}\cdot\vec{k}) \delta^{al} \frac{\partial}{\partial k^a}\frac{\partial}{\partial k^l}-4k^a \frac{\partial}{\partial k^a} b^l\frac{\partial}{\partial k^l}\right]\langle \tilde{v}_i(\vec{k})\tilde{v}_j(-\vec{k})\rangle' \right. \\
\left.-2\left(b^a \frac{\partial}{\partial k^i}-b_i \frac{\partial}{\partial k_{a}}\right) \langle \tilde{v}_a(\vec{k}) \tilde{v}_j(-\vec{k})\rangle'-2\left(b^a \frac{\partial}{\partial k^j}-b_j \frac{\partial}{\partial k_{a}}\right)\langle \tilde{v}_i (\vec{k}) \tilde{v}_a(-\vec{k})\rangle' \right\}\;.
\end{multline}
and conclude that if the two point correlator is invariant under SCT it must satisfy:
\begin{empheq}[box=\fbox]{align}
\left[4 (\Delta_v-3)  (b^l \frac{\partial}{\partial k^l})+2(\vec{b}\cdot\vec{k}) \delta^{al} \frac{\partial}{\partial k^a}\frac{\partial}{\partial k^l}-4k^a \frac{\partial}{\partial k^a} b^l\frac{\partial}{\partial k^l}\right]\langle \tilde{v}_i(\vec{k})\tilde{v}_j(-\vec{k})\rangle' &\no\\
-2\left(b^a \frac{\partial}{\partial k^i}-b_i \frac{\partial}{\partial k_{a}}\right) \langle \tilde{v}_a(\vec{k}) \tilde{v}_j(-\vec{k})\rangle'-2\left(b^a \frac{\partial}{\partial k^j}-b_j \frac{\partial}{\partial k_{a}}\right)\langle \tilde{v}_i (\vec{k}) \tilde{v}_a(-\vec{k})\rangle'=&\;0\;.
\end{empheq}

\subsubsection{SCT 3-point function}
%%%%%%%%%%%%%%%%%%%%%%%%%%%%
The procedure to derive the Ward identity for SCT for the 3-point correlator is the same as in Sec. \ref{scttp} for the 2-point correlator until Eq. (\ref{wardsctftol}). After (\ref{wardsctftol}), we need to see the action of the operators on the delta function. It easy to check that if the correlator satisfy the dilatation Ward identity and 
\begin{multline}
\sum_{n=1}^{3}\left(k_{(n)l}\frac{\partial}{\partial k^{a}_n}-k_{(n)a}\frac{\partial}{\partial k_{(n)}^l}\right)\langle \phi (\vec{k}_1)  v_i (\vec{k}_2) v_j (\vec{k}_3) \rangle' \\
-\delta_{ia}\langle \phi (\vec{k}_1)  v_l (\vec{k}_2) v_j (\vec{k}_3) \rangle'+\delta_{il}\langle \phi (\vec{k}_1)  v_a (\vec{k}_2) v_j (\vec{k}_3) \rangle'\\
-\delta_{aj}\langle \phi (\vec{k}_1)  v_i (\vec{k}_2) v_l (\vec{k}_3) \rangle'+\delta_{lj}\langle \phi (\vec{k}_1)  v_i (\vec{k}_2) v_a (\vec{k}_3) \rangle'=0\;,
\end{multline}
the delta function is not affected by the operators. The later equation is related to rotation covariance and spin invariance \cite{Maldacena:2011nz}. Taking this into account, the corresponding equation (\ref{wardsctftodel}) is
{\small
\begin{multline}
\int \frac{Dk}{(2 \pi)^\frac{9}{2}}e^{i\vec{k}_n\cdot\vec{x}_n}\delta(\vec{k}_{123}) \delta_{\vec{b}}\langle \phi (\vec{k}_1)  v_i (\vec{k}_2) v_j (\vec{k}_3) \rangle' =\\ 
\int \frac{Dk}{(2 \pi)^\frac{9}{2}}e^{i\vec{k}_n\cdot\vec{x}_n}\delta(\vec{k}_{123})\left\{\sum_{a=1}^3[2(\Delta_v-3)(\vec{b}\cdot\vec{\partial}_{a})+(\vec{b}\cdot\vec{k}_a)\nabla_a^2-2(\vec{k}_a \cdot \vec{\partial}_a)(\vec{b}\cdot \vec{\partial}_a)]\langle \phi (\vec{k}_1)  v_i (\vec{k}_2) v_j (\vec{k}_3) \rangle'\right.\\ 
\left.-2[b^l\partial_{2i}-b_i\partial_{2l}]\langle \phi (\vec{k}_1)  v_l (\vec{k}_2) v_j (\vec{k}_3) \rangle'-2[b^l\partial_{3j}-b_j\partial_{3l}]\langle \phi (\vec{k}_1)  v_i (\vec{k}_2) v_l (\vec{k}_3) \rangle'\right\},
\end{multline}
}
where $Dk=d^3k_1d^3k_2d^3k_3$ and $n=1,2,3$. Finally, we can see that if the 3-point function is invariant under SCT, then the correlator must satisfy
\begin{empheq}[box=\fbox]{align}
\sum_{a=1}^3\left[2(\Delta_a-3) {\partial}_{k_a^n}+ {k}^n_a \nabla_a^2-2 \vec{k}_a \cdot \vec{\partial}_a {\partial}_{k_a^n} \right]\langle \phi (\vec{k}_1)  v_i (\vec{k}_2) v_j (\vec{k}_3) \rangle' & \no\\
-2[ \delta_{ln}\partial_{2i}-  \delta_{ni}\partial_{2l}]\langle \phi (\vec{k}_1)  v_l (\vec{k}_2) v_j (\vec{k}_3) \rangle'-2[  \delta_{ln}\partial_{3j}- \delta_{jn}\partial_{3l}]\langle \phi (\vec{k}_1)  v_i (\vec{k}_2) v_l (\vec{k}_3) \rangle'&=\; 0. 
\end{empheq}
\section{Some useful identities and properties}\label{AB}
%%%%%%%%%%%%%%%%%%%%%%%%%%%%%%%
%%%%%%%%%%%%%%
\subsection{Derivatives of the two point function}\label{AB1}
%%%%%%%%%%%%%%
The following identities are useful to evaluate the Ward identities for the two point function:
\ba
k_{a}\Delta_{ab} &=& k_{a}\Pi_{ab} = 0, \quad \Delta_{ab}\Delta_{bc}=\Delta_{ac}, \quad \Delta_{aa} = 2\\
\partial_{a}f(k) &=& f'(k)\hat{k}_a, \quad \partial_a \hat{k}_{b} =\frac{1}{k} \Delta_{ab} ,\quad \partial_a \Delta_{bc} = -\frac{1}{k} (\Delta_{ab} \hat{k}_{c} + \Delta_{ac} \hat{k}_{b} )\\
k_{a} \partial_a \Delta_{bc} &=&0, \quad k_{b} \partial_a \Delta_{bc} = -\Delta_{ac}, %\\
%\partial_{a}\Pi_{bc}  &=& -\frac{1}{k} (\Delta_{ab} \hat{k}_{c} + \Delta_{ac} \hat{k}_{b} ) + i k \beta' \epsilon_{bcd} \hat{k}_{a}\hat{k}_{d} + i\beta \epsilon_{bca} \\
%&=& -\frac{1}{k} (\Delta_{ab} \hat{k}_{c} + \Delta_{ac} \hat{k}_{b} ) + i\beta \epsilon_{bcd}\Delta_{ad} + i \frac{d (k\beta)}{dk}  \epsilon_{bcd} \hat{k}_{a}\hat{k}_{d} \\
%k_{a}\partial_{a}\Pi_{bc}  &=& i k \frac{d (k\beta)}{dk}   \epsilon_{bcd} \hat{k}_{d}, \quad k_{b}\partial_{a}\Pi_{bc} = -\Delta_{ac} + i \beta \epsilon_{bca} k_b\\
%\partial_{a}P_{bc}  &=& \frac{dP}{dk} \Delta_{bc}\hat{k}_i-\frac{P}{k} (\Delta_{ab} \hat{k}_{c} + \Delta_{ac} \hat{k}_{b} ) + i \frac{d(k P\beta)}{dk}  \epsilon_{bcd} \hat{k}_{a}\hat{k}_{d} + i P\beta \epsilon_{bcd}\Delta_{da} \\
\ea
For the calculation of the Ward identity of SCT we need the first and second derivative of the two point function. Using the form of the power spectrum (\ref{psab}) we find:
\ba \label{1std}
 \partial_{i}P_{lm}  &=& p \frac{\hat{k}_{i} }{ k }P_{lm} + \frac{k^p}{k} \left[-\alpha (\Delta_{il} \hat{k}_{m} + \Delta_{im} \hat{k}_{l} ) + i\beta \epsilon_{lmc} \Delta_{ic}   \right] \\ \label{2ndd}
  \partial_{j} \partial_{i}P_{lm}  &=& (p-1)\left( p \frac{\hat{k}_{j} \hat{k}_{i} }{ k^2 } P_{lm} + \frac{ \hat{k}_{j} }{k^2} k^p \left[-\alpha (\Delta_{il} \hat{k}_{m} + \Delta_{im} \hat{k}_{l} ) + i\beta \epsilon_{lmc} \Delta_{ic}   \right]  \right) \\ \nonumber
&+& \alpha \frac{k^p}{k^2}  \left[ p  \Delta_{ji}  \Delta_{lm} -  \Delta_{jm}  \Delta_{il} -  \Delta_{jl}  \Delta_{mi} + 2 \hat{k}_{l} \hat{k}_{m} \Delta_{ji}    -(p-1) \left( \hat{k}_{i}\hat{k}_{m} \Delta_{jl} + \hat{k}_{i}\hat{k}_{l} \Delta_{jm} \right)  \right] \\ \nonumber
&+& i\beta (p-1) \frac{k^p}{k^2}  \epsilon_{lmc} \left[  \left( \Delta_{ji} \hat{k}_{c} + \Delta_{jc} \hat{k}_{i}  \right)  \right] .
\ea
With this expressions we find the different contractions of the first and second derivatives:
\ba\label{1stcnt}
\partial_i P_{im} &=& \partial_i P_{mi} = - 2\alpha \frac{k^p}{k}\hat{k}_{m}. \\
%\nabla^2 P_{lm} =  \partial_{i} \partial_{i}P_{lm}  &=& p(p-1)  \frac{ 1 }{ k^2 } P_{lm}   + 2 (p-1) \left( \alpha \frac{k^p}{k^2} \Delta_{lm} + i\beta  \frac{k^p}{k^2}  \epsilon_{lmc}   \hat{k}_{c} \right)+ \alpha \frac{k^p}{k^2}   4 \hat{k}_{l} \hat{k}_{m}   \\  \nonumber
%&=& p(p-1)  \frac{ 1 }{ k^2 } P_{lm}   +   2 (p-1) \frac{1}{k^2} P_{lm}  +  \alpha \frac{k^p}{k^2}   4 \hat{k}_{l} \hat{k}_{m} \\ 
\label{2ndcnt1}
\nabla^2 P_{lm}&=& (p+2)(p-1)  \frac{ 1 }{ k^2 } P_{lm}    +  \alpha \frac{k^p}{k^2}   4 \hat{k}_{l} \hat{k}_{m} \\ \label{2ndcnt2}
\vec{k}\cdot \nabla  \partial_{i} P_{lm} &=&    (p-1) p \frac{ \hat{k}_{i} }{ k } P_{lm} + (p-1)  \frac{ 1 }{ k } k^p \left[-\alpha (\Delta_{il} \hat{k}_{m} + \Delta_{im} \hat{k}_{l} ) + i\beta \epsilon_{lmc} \Delta_{ic}   \right] . 
\ea
%%%%%%%%%%%%%%%%%%%%%%
\subsection{Derivatives of the 3-point function}\label{apdev3p}
%%%%%%%%%%%%%%%%%%%%%%%%%
Here we compute the derivatives of the matrices $T_{ij}^{(n)}$ with respect to $k_{2}$. The derivatives with respect to $k_{3}$ are obtained by changing $k_2 \leftrightarrow k_3$ and $i \leftrightarrow j$  whenever the indices are out of the matrices:
\begin{align}
\partial_{(2)l} T_{ij}^{(1)}&=\frac{\partial}{\partial k_2^l} T_{ij}^{(1)}=-\frac{1}{k_2}\left[\hat{k}_{2i}T_{lj}^{(1)}+\hat{k}_{2a}\Delta_{(3)aj}\Delta_{(2)il}\right]\;,\\
\partial_{(2)l} T_{ij}^{(2)}&=(\hat{k}_{3a}\delta_{ij}-\hat{k}_{3i}\delta_{aj})\frac{\Delta_{(2)la}}{k_2}\;,\\
\partial_{(2)l} T_{ij}^{(3)}&=\frac{\Delta_{(2)al}}{k_2}\left[ -\epsilon_{ija}+\delta_{ia}\epsilon_{jcb}\hat{k}_{2c}\hat{k}_{3b}+(\hat{k}_{2i}\epsilon_{jab}-\hat{k}_{3j}\epsilon_{iab})\hat{k}_{3b}\right]\;,
\end{align} 
here we notices that $k_{2a} \partial_{(2)a}T_{ij}^{(n)}=0$, since $k_{2i}\Delta_{(2)ia}=0$. 
The 2nd derivatives of $T_{ij}^{(n)}$  are:
\begin{align}
\partial_{(2)m}\partial_{(2)l} T_{ij}^{(1)}&=-\frac{\hat{k}_{2m}}{k_2}\partial_{(2)l}T_{ij}^{(1)}-\frac{1}{k_2}\left[\frac{\Delta_{(2)mi}}{k_2}T_{lj}^{(1)}+\frac{\Delta_{(2)il}}{k_2}T_{mj}^{(1)}+\hat{k}_{2a}\Delta_{(3)aj}\partial_{(2)m}\Delta_{(2)il}\right]\;,\\
\partial_{(2)m}\partial_{(2)l} T_{ij}^{(2)}&=(\hat{k}_{3a}\delta_{ij}-\hat{k}_{3i}\delta_{aj})\left[\frac{\partial_{(2)m}\Delta_{(2)la}}{k_2}-\frac{\hat{k}_{2m}\Delta_{(2)la}}{k_2^2}\right]\;,\\
\partial_{(2)m}\partial_{(2)l} T_{ij}^{(3)}&=\left(\frac{\partial_{(2)m}\Delta_{(2)al}}{k_2}-\frac{\hat{k}_{2m}\Delta_{(2)al}}{k_2^2}\right)\left[ -\epsilon_{ija}+\delta_{ia}\epsilon_{jcb}\hat{k}_{2c}\hat{k}_{3b}+(\hat{k}_{2i}\epsilon_{jab}-\hat{k}_{3j}\epsilon_{iab})\hat{k}_{3b}\right]\;,\nonumber \\
&+\frac{\Delta_{(2)al}}{k_2^2}\left[\delta_{ia}\epsilon_{jcb}\Delta_{(2)cm}\hat{k}_{3b}+\Delta_{(2)mi}\epsilon_{jab}\hat{k}_{3b}\right]\;.
\end{align}
Finally, their contractions are:
\begin{align}
\partial_{(2)m}\partial_{(2)m} T_{ij}^{(1)}&=-\frac{2}{k_2^2}\left[T_{ij}^{(1)}-\hat{k}_{2i}\hat{k}_{2a}\Delta_{(3)aj}\right]\;,\\
\partial_{(2)m}\partial_{(2)m} T_{ij}^{(2)}&=-2\frac{T_{ij}^{(2)}}{k_2^2}\;,\\
\partial_{(2)m}\partial_{(2)m} T_{ij}^{(3)}&=-\frac{2}{k_2^2}\left[ T_{ij}^{(3)}+2\hat{k}_{2i}\epsilon_{jab}\hat{k}_{2a}\hat{k}_{3b}\right]\;,
\end{align}
and
\begin{align}
k_{2m}\partial_{(2)m}\partial_{(2)l} T_{ij}^{(m)}&=-\partial_{(2)l}T_{ij}^{(m)}.
\end{align}

%%%%%%%%%%%%%%
\subsection{A useful identity}\label{AB2}
%%%%%%%%%%%%%%
The following identity results useful for dealing with the antisymmetric part of the correlators:
\be\label{ide}
\epsilon_{mli} = (\epsilon_{mci} \hat{k}_{l} + \epsilon_{cli} \hat{k}_{m} + \epsilon_{mlc} \hat{k}_{i}) \hat{k}_{c}.
\ee
This identity can be checked by contracting with $\hat{k}_{i}, \hat{k}_{l} $ and $\hat{k}_{m}$ or by directly calculating the components for $m, l, i = \{1, 2, 3\}$. This identity holds for any unitary vector $\hat{k}$. 
Contracting with the vectors $\hat{k}_2$ and $\hat{k}_3$ we get:
\ba 
\hat{k}_{2m}\epsilon_{mli} &=& (\hat{k}_{2m}\epsilon_{mci} \hat{k}_{3l} + \epsilon_{cli} (\hat{k}_{2}\cdot\hat{k}_{3}) + \hat{k}_{2m}\epsilon_{mlc} \hat{k}_{3i}) \hat{k}_{3c}\\
\hat{k}_{3m}\epsilon_{mli} &=& (\hat{k}_{3m}\epsilon_{mci} \hat{k}_{2l} + \epsilon_{cli} (\hat{k}_{2}\cdot\hat{k}_{3}) + \hat{k}_{3m}\epsilon_{mlc} \hat{k}_{2i}) \hat{k}_{2c}
\ea
which after subtracting and after some rearranging allows us to see that $T^{(3)}_{ij} = T^{(4)}_{ij}$.

%%%%%%%%%%%%%

%\begin{thebibliography}{30}
\bibliographystyle{JHEP} 
\bibliography{Biblio_SVC} 

%%%%%%%%%%%%%
%Non Abelian Gauge fields references
%%%%%%%%%%%%%

%\end{thebibliography}  

\end{document}